# Indirect measurement of $\sin^2\theta_W$ $(M_W)$ using $e^+e^-$ pairs in the $Z$-boson region with $p\bar{p}$ collisions at a center-of-momentum energy of 1.96 TeV

T. Aaltonen,[21] S. Amerio[jj],[39] D. Amidei,[31] A. Anastassov[v],[15] A. Annovi,[17] J. Antos,[12] G. Apollinari,[15] J.A. Appel,[15] T. Arisawa,[52] A. Artikov,[13] J. Asaadi,[47] W. Ashmanskas,[15] B. Auerbach,[2] A. Aurisano,[47] F. Azfar,[38] W. Badgett,[15] T. Bae,[25] A. Barbaro-Galtieri,[26] V.E. Barnes,[43] B.A. Barnett,[23] P. Barria[ll],[41] P. Bartos,[12] M. Bauce[jj],[39] F. Bedeschi,[41] S. Behari,[15] G. Bellettini[kk],[41] J. Bellinger,[54] D. Benjamin,[14] A. Beretvas,[15] A. Bhatti,[45] K.R. Bland,[5] B. Blumenfeld,[23] A. Bocci,[14] A. Bodek,[44] D. Bortoletto,[43] J. Boudreau,[42] A. Boveia,[11] L. Brigliadori[ii],[6] C. Bromberg,[32] E. Brucken,[21] J. Budagov,[13] H.S. Budd,[44] K. Burkett,[15] G. Busetto[jj],[39] P. Bussey,[19] P. Butti[kk],[41] A. Buzatu,[19] A. Calamba,[10] S. Camarda,[4] M. Campanelli,[28] F. Canelli[cc],[11] B. Carls,[22] D. Carlsmith,[54] R. Carosi,[41] S. Carrillo[l],[16] B. Casal[j],[9] M. Casarsa,[48] A. Castro[ii],[6] P. Catastini,[20] D. Cauz[qqrr],[48] V. Cavaliere,[22] M. Cavalli-Sforza,[4] A. Cerri[e],[26] L. Cerrito[q],[28] Y.C. Chen,[1] M. Chertok,[7] G. Chiarelli,[41] G. Chlachidze,[15] K. Cho,[25] D. Chokheli,[13] A. Clark,[18] C. Clarke,[53] M.E. Convery,[15] J. Conway,[7] M. Corbo[y],[15] M. Cordelli,[17] C.A. Cox,[7] D.J. Cox,[7] M. Cremonesi,[41] D. Cruz,[47] J. Cuevas[x],[9] R. Culbertson,[15] N. d'Ascenzo[u],[15] M. Datta[ff],[15] P. de Barbaro,[44] L. Demortier,[45] M. Deninno,[6] M. D'Errico[jj],[39] F. Devoto,[21] A. Di Canto[kk],[41] B. Di Ruzza[p],[15] J.R. Dittmann,[5] S. Donati[kk],[41] M. D'Onofrio,[27] M. Dorigo[ss],[48] A. Driutti[qqrr],[48] K. Ebina,[52] R. Edgar,[31] A. Elagin,[47] R. Erbacher,[7] S. Errede,[22] B. Esham,[22] S. Farrington,[38] J.P. Fernández Ramos,[29] R. Field,[16] G. Flanagan[s],[15] R. Forrest,[7] M. Franklin,[20] J.C. Freeman,[15] H. Frisch,[11] Y. Funakoshi,[52] C. Galloni[kk],[41] A.F. Garfinkel,[43] P. Garosi[ll],[41] H. Gerberich,[22] E. Gerchtein,[15] S. Giagu,[46] V. Giakoumopoulou,[3] K. Gibson,[42] C.M. Ginsburg,[15] N. Giokaris,[3] P. Giromini,[17] G. Giurgiu,[23] V. Glagolev,[13] D. Glenzinski,[15] M. Gold,[34] D. Goldin,[47] A. Golossanov,[15] G. Gomez,[9] G. Gomez-Ceballos,[30] M. Goncharov,[30] O. González López,[29] I. Gorelov,[34] A.T. Goshaw,[14] K. Goulianos,[45] E. Gramellini,[6] S. Grinstein,[4] C. Grosso-Pilcher,[11] R.C. Group,[51,15] J. Guimaraes da Costa,[20] S.R. Hahn,[15] J.Y. Han,[44] F. Happacher,[17] K. Hara,[49] M. Hare,[50] R.F. Harr,[53] T. Harrington-Taber[m],[15] K. Hatakeyama,[5] C. Hays,[38] J. Heinrich,[40] M. Herndon,[54] A. Hocker,[15] Z. Hong,[47] W. Hopkins[f],[15] S. Hou,[1] R.E. Hughes,[35] U. Husemann,[55] M. Hussein[aa],[32] J. Huston,[32] G. Introzzi[nnoo],[41] M. Iori[pp],[46] A. Ivanov[o],[7] E. James,[15] D. Jang,[10] B. Jayatilaka,[15] E.J. Jeon,[25] S. Jindariani,[15] M. Jones,[43] K.K. Joo,[25] S.Y. Jun,[10] T.R. Junk,[15] M. Kambeitz,[24] T. Kamon,[25,47] P.E. Karchin,[53] A. Kasmi,[5] Y. Kato[n],[37] W. Ketchum[gg],[11] J. Keung,[40] B. Kilminster[cc],[15] D.H. Kim,[25] H.S. Kim,[25] J.E. Kim,[25] M.J. Kim,[17] S.H. Kim,[49] S.B. Kim,[25] Y.J. Kim,[25] Y.K. Kim,[11] N. Kimura,[52] M. Kirby,[15] K. Knoepfel,[15] K. Kondo,[52,*] D.J. Kong,[25] J. Konigsberg,[16] A.V. Kotwal,[14] M. Kreps,[24] J. Kroll,[40] M. Kruse,[14] T. Kuhr,[24] M. Kurata,[49] A.T. Laasanen,[43] S. Lammel,[15] M. Lancaster,[28] K. Lannon[w],[35] G. Latino[ll],[41] H.S. Lee,[25] J.S. Lee,[25] S. Leo,[41] S. Leone,[41] J.D. Lewis,[15] A. Limosani[r],[14] E. Lipeles,[40] A. Lister[a],[18] H. Liu,[51] Q. Liu,[43] T. Liu,[15] S. Lockwitz,[55] A. Loginov,[55] D. Lucchesi[jj],[39] A. Lucà,[17] J. Lueck,[24] P. Lujan,[26] P. Lukens,[15] G. Lungu,[45] J. Lys,[26] R. Lysak[d],[12] R. Madrak,[15] P. Maestro[ll],[41] S. Malik,[45] G. Manca[b],[27] A. Manousakis-Katsikakis,[3] L. Marchese[hh],[6] F. Margaroli,[46] P. Marino[mm],[41] M. Martínez,[4] K. Matera,[22] M.E. Mattson,[53] A. Mazzacane,[15] P. Mazzanti,[6] R. McNulty[i],[27] A. Mehta,[27] P. Mehtala,[21] C. Mesropian,[45] T. Miao,[15] D. Mietlicki,[31] A. Mitra,[1] H. Miyake,[49] S. Moed,[15] N. Moggi,[6] C.S. Moon[y],[15] R. Moore[ddee],[15] M.J. Morello[mm],[41] A. Mukherjee,[15] Th. Muller,[24] P. Murat,[15] M. Mussini[ii],[6] J. Nachtman[m],[15] Y. Nagai,[49] J. Naganoma,[52] I. Nakano,[36] A. Napier,[50] J. Nett,[47] C. Neu,[51] T. Nigmanov,[42] L. Nodulman,[2] S.Y. Noh,[25] O. Norniella,[22] L. Oakes,[38] S.H. Oh,[14] Y.D. Oh,[25] I. Oksuzian,[51] T. Okusawa,[37] R. Orava,[21] L. Ortolan,[4] C. Pagliarone,[48] E. Palencia[e],[9] P. Palni,[34] V. Papadimitriou,[15] W. Parker,[54] G. Pauletta[qqrr],[48] M. Paulini,[10] C. Paus,[30] T.J. Phillips,[14] G. Piacentino,[41] E. Pianori,[40] J. Pilot,[7] K. Pitts,[22] C. Plager,[8] L. Pondrom,[54] S. Poprocki[f],[15] K. Potamianos,[26] A. Pranko,[26] F. Prokoshin[z],[13] F. Ptohos[g],[17] G. Punzi[kk],[41] N. Ranjan,[43] I. Redondo Fernández,[29] P. Renton,[38] M. Rescigno,[46] F. Rimondi,[6,*] L. Ristori,[41,15] A. Robson,[19] T. Rodriguez,[40] S. Rolli[h],[50] M. Ronzani[kk],[41] R. Roser,[15] J.L. Rosner,[11] F. Ruffini[ll],[41] A. Ruiz,[9] J. Russ,[10] V. Rusu,[15] W.K. Sakumoto,[44] Y. Sakurai,[52] L. Santi[qqrr],[48] K. Sato,[49] V. Saveliev[u],[15] A. Savoy-Navarro[y],[15] P. Schlabach,[15] E.E. Schmidt,[15] T. Schwarz,[31] L. Scodellaro,[9] F. Scuri,[41] S. Seidel,[34] Y. Seiya,[37] A. Semenov,[13] F. Sforza[kk],[41] S.Z. Shalhout,[7] T. Shears,[27] P.F. Shepard,[42] M. Shimojima[t],[49] M. Shochet,[11] I. Shreyber-Tecker,[33] A. Simonenko,[13] K. Sliwa,[50] J.R. Smith,[7] F.D. Snider,[15] H. Song,[42] V. Sorin,[4] R. St. Denis,[19] M. Stancari,[15] D. Stentz[v],[15] J. Strologas,[34] Y. Sudo,[49] A. Sukhanov,[15] I. Suslov,[13] K. Takemasa,[49] Y. Takeuchi,[49] J. Tang,[11] M. Tecchio,[31] P.K. Teng,[1] J. Thom[f],[15] E. Thomson,[40] V. Thukral,[47] D. Toback,[47] S. Tokar,[12] K. Tollefson,[32] T. Tomura,[49] D. Tonelli[e],[15] S. Torre,[17] D. Torretta,[15] P. Totaro,[39] M. Trovato[mm],[41] F. Ukegawa,[49] S. Uozumi,[25] F. Vázquez[l],[16] G. Velev,[15] C. Vellidis,[15] C. Vernieri[mm],[41] M. Vidal,[43] R. Vilar,[9] J. Vizán[bb],[9] M. Vogel,[34] G. Volpi,[17] P. Wagner,[40] R. Wallny[j],[15] S.M. Wang,[1] D. Waters,[28]

W.C. Wester III,[15] D. Whiteson[c],[40] A.B. Wicklund,[2] S. Wilbur,[7] H.H. Williams,[40] J.S. Wilson,[31] P. Wilson,[15] B.L. Winer,[35] P. Wittich[f],[15] S. Wolbers,[15] H. Wolfe,[35] T. Wright,[31] X. Wu,[18] Z. Wu,[5] K. Yamamoto,[37] D. Yamato,[37] T. Yang,[15] U.K. Yang,[25] Y.C. Yang,[25] W.-M. Yao,[26] G.P. Yeh,[15] K. Yi[m],[15] J. Yoh,[15] K. Yorita,[52] T. Yoshida[k],[37] G.B. Yu,[14] I. Yu,[25] A.M. Zanetti,[48] Y. Zeng,[14] C. Zhou,[14] and S. Zucchelli[ii6]

(CDF Collaboration)[†]

[1]*Institute of Physics, Academia Sinica, Taipei, Taiwan 11529, Republic of China*
[2]*Argonne National Laboratory, Argonne, Illinois 60439, USA*
[3]*University of Athens, 157 71 Athens, Greece*
[4]*Institut de Fisica d'Altes Energies, ICREA, Universitat Autonoma de Barcelona, E-08193, Bellaterra (Barcelona), Spain*
[5]*Baylor University, Waco, Texas 76798, USA*
[6]*Istituto Nazionale di Fisica Nucleare Bologna, [ii]University of Bologna, I-40127 Bologna, Italy*
[7]*University of California, Davis, Davis, California 95616, USA*
[8]*University of California, Los Angeles, Los Angeles, California 90024, USA*
[9]*Instituto de Fisica de Cantabria, CSIC-University of Cantabria, 39005 Santander, Spain*
[10]*Carnegie Mellon University, Pittsburgh, Pennsylvania 15213, USA*
[11]*Enrico Fermi Institute, University of Chicago, Chicago, Illinois 60637, USA*
[12]*Comenius University, 842 48 Bratislava, Slovakia; Institute of Experimental Physics, 040 01 Kosice, Slovakia*
[13]*Joint Institute for Nuclear Research, RU-141980 Dubna, Russia*
[14]*Duke University, Durham, North Carolina 27708, USA*
[15]*Fermi National Accelerator Laboratory, Batavia, Illinois 60510, USA*
[16]*University of Florida, Gainesville, Florida 32611, USA*
[17]*Laboratori Nazionali di Frascati, Istituto Nazionale di Fisica Nucleare, I-00044 Frascati, Italy*
[18]*University of Geneva, CH-1211 Geneva 4, Switzerland*
[19]*Glasgow University, Glasgow G12 8QQ, United Kingdom*
[20]*Harvard University, Cambridge, Massachusetts 02138, USA*
[21]*Division of High Energy Physics, Department of Physics, University of Helsinki, FIN-00014, Helsinki, Finland; Helsinki Institute of Physics, FIN-00014, Helsinki, Finland*
[22]*University of Illinois, Urbana, Illinois 61801, USA*
[23]*The Johns Hopkins University, Baltimore, Maryland 21218, USA*
[24]*Institut für Experimentelle Kernphysik, Karlsruhe Institute of Technology, D-76131 Karlsruhe, Germany*
[25]*Center for High Energy Physics: Kyungpook National University, Daegu 702-701, Korea; Seoul National University, Seoul 151-742, Korea; Sungkyunkwan University, Suwon 440-746, Korea; Korea Institute of Science and Technology Information, Daejeon 305-806, Korea; Chonnam National University, Gwangju 500-757, Korea; Chonbuk National University, Jeonju 561-756, Korea; Ewha Womans University, Seoul, 120-750, Korea*
[26]*Ernest Orlando Lawrence Berkeley National Laboratory, Berkeley, California 94720, USA*
[27]*University of Liverpool, Liverpool L69 7ZE, United Kingdom*
[28]*University College London, London WC1E 6BT, United Kingdom*
[29]*Centro de Investigaciones Energeticas Medioambientales y Tecnologicas, E-28040 Madrid, Spain*
[30]*Massachusetts Institute of Technology, Cambridge, Massachusetts 02139, USA*
[31]*University of Michigan, Ann Arbor, Michigan 48109, USA*
[32]*Michigan State University, East Lansing, Michigan 48824, USA*
[33]*Institution for Theoretical and Experimental Physics, ITEP, Moscow 117259, Russia*
[34]*University of New Mexico, Albuquerque, New Mexico 87131, USA*
[35]*The Ohio State University, Columbus, Ohio 43210, USA*
[36]*Okayama University, Okayama 700-8530, Japan*
[37]*Osaka City University, Osaka 558-8585, Japan*
[38]*University of Oxford, Oxford OX1 3RH, United Kingdom*
[39]*Istituto Nazionale di Fisica Nucleare, Sezione di Padova, [jj]University of Padova, I-35131 Padova, Italy*
[40]*University of Pennsylvania, Philadelphia, Pennsylvania 19104, USA*
[41]*Istituto Nazionale di Fisica Nucleare Pisa, [kk]University of Pisa, [ll]University of Siena, [mm]Scuola Normale Superiore, I-56127 Pisa, Italy, [nn]INFN Pavia, I-27100 Pavia, Italy, [oo]University of Pavia, I-27100 Pavia, Italy*
[42]*University of Pittsburgh, Pittsburgh, Pennsylvania 15260, USA*
[43]*Purdue University, West Lafayette, Indiana 47907, USA*
[44]*University of Rochester, Rochester, New York 14627, USA*
[45]*The Rockefeller University, New York, New York 10065, USA*
[46]*Istituto Nazionale di Fisica Nucleare, Sezione di Roma 1, [pp]Sapienza Università di Roma, I-00185 Roma, Italy*

[47]Mitchell Institute for Fundamental Physics and Astronomy, Texas A&M University, College Station, Texas 77843, USA
[48]Istituto Nazionale di Fisica Nucleare Trieste, [qq]Gruppo Collegato di Udine, [rr]University of Udine, I-33100 Udine, Italy, [ss]University of Trieste, I-34127 Trieste, Italy
[49]University of Tsukuba, Tsukuba, Ibaraki 305, Japan
[50]Tufts University, Medford, Massachusetts 02155, USA
[51]University of Virginia, Charlottesville, Virginia 22906, USA
[52]Waseda University, Tokyo 169, Japan
[53]Wayne State University, Detroit, Michigan 48201, USA
[54]University of Wisconsin, Madison, Wisconsin 53706, USA
[55]Yale University, New Haven, Connecticut 06520, USA

Drell-Yan lepton pairs are produced in the process $p\bar{p} \to e^+e^- + X$ through an intermediate $\gamma^*/Z$ boson. The lepton angular distributions are used to provide information on the electroweak-mixing parameter $\sin^2\theta_W$ via its observable effective-leptonic $\sin^2\theta_W$, or $\sin^2\theta^{\rm lept}_{\rm eff}$. A new method to infer $\sin^2\theta_W$, or equivalently, the the $W$-boson mass $M_W$ in the on-shell scheme, is developed and tested using a previous CDF Run II measurement of angular distributions from electron pairs in a sample corresponding to 2.1 fb$^{-1}$ of integrated luminosity from $p\bar{p}$ collisions at a center-of-momentum energy of 1.96 TeV. The value of $\sin^2\theta^{\rm lept}_{\rm eff}$ is found to be $0.2328 \pm 0.0010$. Within a specified context of the standard model, this results in $\sin^2\theta_W = 0.2246 \pm 0.0009$ which corresponds to a $W$-boson mass of $80.297 \pm 0.048$ GeV/$c^2$, in agreement with previous determinations in electron-position collisions and at the Tevatron collider.

## I. INTRODUCTION

The angular distribution of electrons from the Drell-Yan [1] process is used to measure the electroweak-mixing parameter $\sin^2\theta_W$ [2]. At the Tevatron, Drell-Yan pairs are produced by the process $p\bar{p} \to e^+e^- + X$, where the $e^+e^-$ pair is produced through an intermediate $\gamma^*/Z$ boson, and $X$ is the hadronic final state associated with the production of the boson. In the standard model, the Drell-Yan process at the Born level is described by two parton-level amplitudes:

$$q\bar{q} \to \gamma^* \to e^+e^-, \text{ and}$$
$$q\bar{q} \to Z \to e^+e^-.$$

The fermions ($f$) couple to the virtual photon via a vector coupling, $Q_f\gamma_\mu$, where $Q_f$ is the fermion charge (in units of $e$). The fermion coupling to $Z$ bosons consists of both vector ($V$) and axial-vector ($A$) couplings: $g^f_V\gamma_\mu + g^f_A\gamma_\mu\gamma_5$. The Born-level couplings are

$$g^f_V = T^f_3 - 2Q_f \sin^2\theta_W$$
$$g^f_A = T^f_3,$$

where $T^f_3$ is the third component of the fermion weak isospin. The $\sin^2\theta_W$ parameter is related to the $W$-boson mass $M_W$, and the $Z$-boson mass $M_Z$, by the relationship $\sin^2\theta_W = 1 - M^2_W/M^2_Z$ which holds to all orders in the on-shell scheme. These couplings have been investigated both at the Tevatron [3, 4], and at LEP-1 and SLD [5].

In this article, the parameter $\sin^2\theta_W$ is inferred from a previous measurement [6] of the angular distribution of Drell-Yan $e^+e^-$ pairs produced at the Tevatron. The measurement investigates higher-order quantum chromodynamic (QCD) corrections to the angular distribution,

* Deceased
† With visitors from [a]University of British Columbia, Vancouver, BC V6T 1Z1, Canada, [b]Istituto Nazionale di Fisica Nucleare, Sezione di Cagliari, 09042 Monserrato (Cagliari), Italy, [c]University of California Irvine, Irvine, CA 92697, USA, [d]Institute of Physics, Academy of Sciences of the Czech Republic, 182 21, Czech Republic, [e]CERN, CH-1211 Geneva, Switzerland, [f]Cornell University, Ithaca, NY 14853, USA, [g]University of Cyprus, Nicosia CY-1678, Cyprus, [h]Office of Science, U.S. Department of Energy, Washington, DC 20585, USA, [i]University College Dublin, Dublin 4, Ireland, [j]ETH, 8092 Zürich, Switzerland, [k]University of Fukui, Fukui City, Fukui Prefecture, Japan 910-0017, [l]Universidad Iberoamericana, Lomas de Santa Fe, México, C.P. 01219, Distrito Federal, [m]University of Iowa, Iowa City, IA 52242, USA, [n]Kinki University, Higashi-Osaka City, Japan 577-8502, [o]Kansas State University, Manhattan, KS 66506, USA, [p]Brookhaven National Laboratory, Upton, NY 11973, USA, [q]Queen Mary, University of London, London, E1 4NS, United Kingdom, [r]University of Melbourne, Victoria 3010, Australia, [s]Muons, Inc., Batavia, IL 60510, USA, [t]Nagasaki Institute of Applied Science, Nagasaki 851-0193, Japan, [u]National Research Nuclear University, Moscow 115409, Russia, [v]Northwestern University, Evanston, IL 60208, USA, [w]University of Notre Dame, Notre Dame, IN 46556, USA, [x]Universidad de Oviedo, E-33007 Oviedo, Spain, [y]CNRS-IN2P3, Paris, F-75205 France, [z]Universidad Tecnica Federico Santa Maria, 110v Valparaiso, Chile, [aa]The University of Jordan, Amman 11942, Jordan, [bb]Universite catholique de Louvain, 1348 Louvain-La-Neuve, Belgium, [cc]University of Zürich, 8006 Zürich, Switzerland, [dd]Massachusetts General Hospital, Boston, MA 02114 USA, [ee]Harvard Medical School, Boston, MA 02114 USA, [ff]Hampton University, Hampton, VA 23668, USA, [gg]Los Alamos National Laboratory, Los Alamos, NM 87544, USA, [hh]Università degli Studi di Napoli Federico I, I-80138 Napoli, Italy

using electron pairs in the $Z$-boson region 66–116 GeV/$c^2$ from 2.1 fb$^{-1}$ of collisions. This analysis utilizes the results of that measurement to test a new method to obtain $\sin^2\theta_W$. One of the measurements, the $A_4$ angular coefficient, is sensitive to $\sin^2\theta_W$ and is compared with QCD predictions for various values of $\sin^2\theta_W$. The predictions also include electroweak-radiative corrections comparable to those utilized at LEP-1 and SLD [5].

Section II provides an overview of both the electron angular distributions and the method used to obtain $\sin^2\theta_W$. Section III discusses QCD calculations required by the new method. A technique to use and incorporate electroweak radiative-correction form factors for high-energy $e^+e^-$ collisions into the Drell-Yan process is presented. Section IV reviews and documents the event sample, simulation of the data, and methods used in the previous measurement, and describes how the measurement is used in this analysis. Section V describes the systematic uncertainties. Finally, Sec. VI gives the results, and Sec. VII the summary. The units $\hbar = c = 1$ are used for equations and symbols, but standard units are used for numerical values.

## II. ELECTRON ANGULAR DISTRIBUTIONS

The angular distribution of electrons in the boson rest frame is governed by the polarization state of the $\gamma^*/Z$ boson. In amplitudes at higher order than tree level, initial-state QCD interactions of the colliding partons impart transverse momentum, relative to the collision axis, to the $\gamma^*/Z$ boson. This affects the polarization states.

The polar and azimuthal angles of the $e^-$ in the rest frame of the boson are denoted as $\vartheta$ and $\varphi$, respectively. For this analysis, the ideal positive-$z$ axis coincides with the direction of the incoming quark so that $\vartheta$ parallels the definition used in $e^+e^-$ collisions at LEP [5]. This frame is approximated by the Collins-Soper (CS) rest frame [7] for $p\bar{p}$ collisions. The CS frame is reached from the laboratory frame via a Lorentz boost along the laboratory $z$ axis into a frame where the $z$ component of the lepton-pair momentum is zero, followed by a boost along the transverse momentum of the pair. The transverse momentum ($P_{\rm T}$) in a reference frame is the magnitude of momentum transverse to the $z$ axis. Within the CS frame, the $z$ axis for the polar angle is the angular bisector between the proton direction and the negative of the anti-proton direction. The $x$ axis for the azimuthal angle is the direction of the lepton-pair $P_{\rm T}$. At $P_{\rm T} = 0$, the CS and laboratory coordinate systems are the same, and if the incoming quark of the Drell-Yan parton amplitude is from the proton, the $z$ axis and quark directions coincide.

The general structure of the Drell-Yan lepton angular distribution in the boson rest frame consists of nine helicity cross sections [8],

$$
\begin{aligned}
\frac{dN}{d\Omega} \propto\ & (1+\cos^2\vartheta) + \\
& A_0\,\frac{1}{2}\,(1-3\cos^2\vartheta) + \\
& A_1\,\sin 2\vartheta\cos\varphi + \\
& A_2\,\frac{1}{2}\,\sin^2\vartheta\cos 2\varphi + \\
& A_3\,\sin\vartheta\cos\varphi + \\
& A_4\,\cos\vartheta + \\
& A_5\,\sin^2\vartheta\sin 2\varphi + \\
& A_6\,\sin 2\vartheta\sin\varphi + \\
& A_7\,\sin\vartheta\sin\varphi\,.
\end{aligned}
$$

The $A_{0-7}$ coefficients are cross section ratios, and are functions of the boson kinematic variables. They vanish at $P_{\rm T} = 0$, except for the electroweak part of $A_4$ responsible for the forward-backward $e^-$ asymmetry in $\cos\vartheta$. The $A_4$ coefficient is relatively uniform across the range of transverse momentum where the cross section is large, but slowly drops for larger values of $P_{\rm T}$ where the cross section is very small. The $A_{5-7}$ coefficients appear at second order in the QCD strong coupling, $\alpha_s$, and are small in the CS frame [8]. Hereafter, the angles ($\vartheta$, $\varphi$) and the angular coefficients $A_{0-7}$ are specific to the CS rest frame.

The $A_4\cos\vartheta$ term is parity violating, and is due to vector and axial-vector current amplitude interference. Its presence adds an asymmetry to the $\varphi$-integrated $\cos\vartheta$ cross section. Two sources contribute: the interference between the $Z$-boson vector and axial-vector amplitudes, and the interference between the photon vector and $Z$-boson axial-vector amplitudes. The asymmetric component from the $\gamma$-$Z$ interference cross section is proportional to $g_A^f$. The asymmetric component from $Z$ boson self-interference has a coupling factor that is a product of $g_V^f/g_A^f$ from the electron and quark vertices, and thus is related to $\sin^2\theta_W$. At the Born level, this product is

$$(1-4|Q_e|\sin^2\theta_W)\,(1-4|Q_q|\sin^2\theta_W),$$

where $e$ and $q$ denote the electron and quark, respectively. For the Drell-Yan process, the quarks are predominantly light quarks: $u$, $d$, or $s$. As $\sin^2\theta_W \approx 0.223$, the coupling factor has an enhanced sensitivity to $\sin^2\theta_W$ at the electron-$Z$ vertex. A 1% variation in $\sin^2\theta_W$ changes the electron factor (containing $Q_e$) by $\approx 8\%$, while the quark factor (containing $Q_q$) changes by $\approx 1.5\%$ for the $u$ quark, and $\approx 0.4\%$ for the $d$ and $s$ quarks. Loop and vertex electroweak-radiative corrections are multiplicative form-factor corrections to the couplings that change their value by a few percent.

Traditionally, $\sin^2\theta_W$ is inferred from the forward-backward asymmetry of the $e^-$ $\cos\vartheta$ distribution as a function of the dielectron-pair mass. The new method for the inference of $\sin^2\theta_W$ has two inputs: an experimental measurement of the $A_4$ angular-distribution coefficient, and predictions of the $A_4$ coefficient for various input values of $\sin^2\theta_W$. Electroweak and QCD radiative corrections are included in the predictions of the $A_4$ coefficient.

The new method to infer $\sin^2\theta_W$ utilizes the value of the cross-section weighted average, $\bar{A}_4$, for both the experimental input and predictions. The average is

$$\bar{A}_4 = \frac{1}{\sigma}\int_{-\infty}^{\infty} dy \int_0^{\infty} dP_{\rm T}^2 \int dM\ A_4 \frac{d^3\sigma}{dy dP_{\rm T}^2 dM},$$

where $\sigma$ is the integrated cross-section, and $y$, $P_{\rm T}$, and $M$ are the lepton-pair rapidity, transverse momentum, and mass, respectively. The energy and momentum of particles are denoted as $E$ and $P$, respectively. For a given coordinate frame, the rapidity is $y = \frac{1}{2}\ln[\,(E+P_{\rm z})/(E-P_{\rm z})\,]$, where $P_{\rm z}$ is the component of momentum along the $z$ axis of the coordinate frame. The mass integration is limited to the $Z$-boson region 66–116 GeV/$c^2$.

The experimental input for the $\bar{A}_4$ coefficient is derived from a previous measurement of the angular-distribution coefficients $A_0$, $A_2$, $A_3$, and $A_4$, in independent ranges of the dielectron-pair $P_{\rm T}$ [6]. In this analysis, the individual measurements for the $A_4$ coefficient are combined into an average. The predictions provide the relationship between $\sin^2\theta_W$ and $\bar{A}_4$. The QCD predictions of $\bar{A}_4$ include an implementation of electroweak radiative corrections derived from an approach adopted at LEP [9].

## III. ENHANCED QCD PREDICTIONS

Drell-Yan process calculations with QCD radiation do not typically include the full electroweak-radiative corrections. However, the QCD, quantum electrodynamic (QED), and weak corrections can be organized to be individually gauge invariant so that they can be applied separately and independently.

QED radiative corrections with photons in the final state are not included in the calculation of the $\bar{A}_4$ coefficient. Instead, they are applied in the physics and detector simulation of the Drell-Yan process used in the measurement of the $A_4$ coefficients. For the process $q\bar{q} \to e^+e^-$, QED final-state radiation is most important, and is included. The effects of QED radiative corrections are removed from the measurement of the $A_4$ coefficients.

The Drell-Yan process and the production of quark pairs in high energy $e^+e^-$ collisions are analog processes: $q\bar{q} \to e^-e^+$ and $e^-e^+ \to q\bar{q}$. At the Born level, the process amplitudes are of the same form except for the interchange of the electron and quark labels. Electroweak radiative corrections, calculated and extensively used for precision fits of LEP-1 and SLD measurements to the standard model [5], can be applied to the Drell-Yan process.

In the remainder of this section, the technique used to incorporate independently calculated electroweak radiative corrections for $e^+e^-$ collisions into existing QCD calculations for the Drell-Yan process is presented. The results of the QCD calculations for the value of the $\bar{A}_4$ coefficient are also presented.

### A. Electroweak radiative corrections

The effects of electroweak radiative corrections are incorporated into Drell-Yan QCD calculations via form factors for fermion-pair production in $e^+e^-$ collisions, $e^+e^- \to Z \to f\bar{f}$. The form factors are calculated by ZFITTER 6.43 [9], which is used with LEP-1 and SLD measurement inputs for standard-model tests [5]. It is a semi-analytical calculation for fermion-pair production and radiative corrections for high-energy $e^+e^-$ collisions. The set of radiative corrections in each form factor is gauge invariant. Thus it includes $W$-boson loops in the photon propagator and $Z$ propagators at fermion-photon vertices. Consequently, the weak and QED corrections are separately gauge invariant. The renormalization scheme used by ZFITTER is the on-shell scheme [10], where particle masses are on-shell, and

$$\sin^2\theta_W = 1 - M_W^2/M_Z^2 \tag{1}$$

holds to all orders of perturbation theory by definition. Since the $Z$-boson mass is accurately known (to $\pm 0.0021$ GeV/$c^2$ [5]), the inference of $\sin^2\theta_W$ is equivalent to an indirect $W$-boson mass measurement.

Form factors calculated by ZFITTER are stored for later use in QCD calculations. Details of the form-factor calculation with its specific standard-model assumptions and parameters are presented in Appendix A. The calculated form factors are $\rho_{eq}$, $\kappa_e$, $\kappa_q$, and $\kappa_{eq}$, where the label $e$ denotes an electron, and $q$ a quark. As the calculations use the massless-fermion approximation, the form factors only depend on the charge and weak isospin of the fermions. Consequently, the stored form factors are distinguished by three labels: $e$ (electron type), $u$ (up-quark type), and $d$ (down-quark type). The form factors are complex valued, and functions of the $\sin^2\theta_W$ parameter and the Mandelstam $s$ variable of the $e^+e^- \to Z \to f\bar{f}$ process.

The first three form factors can be trivially incorporated into the $q\bar{q} \to Z \to e^+e^-$ interaction currents. The Born-level $g_A^f$ and $g_V^f$ couplings within the currents are replaced with

$$g_V^f \to \sqrt{\rho_{eq}}\,(T_3^f - 2Q_f\kappa_f\,\sin^2\theta_W),\ \text{and}$$
$$g_A^f \to \sqrt{\rho_{eq}}\,T_3^f,$$

where $f = e$ or $q$. The resulting electron-quark current-current interaction amplitude contains a term proportional to $\kappa_e\kappa_q\sin^4\theta_W$. However, as this is an approximation of the desired coefficient, $\kappa_{eq}\sin^4\theta_W$, a further correction to the amplitude (which is discussed in Sec. III B) is required.

The combination $\kappa_f\sin^2\theta_W$, called an effective-mixing parameter, is directly accessible from measurements of the asymmetry in the $\cos\vartheta$ distribution. However, neither the $\sin^2\theta_W$ parameter nor the form factors can be inferrred from experimental measurements without the standard model. The effective-mixing parameters are denoted as $\sin^2\theta_{\rm eff}$ to distinguish them from the on-shell definition of $\sin^2\theta_W$ (Eq. (1)). The Drell-Yan process is most sensitive to the parameter $\sin^2\theta_{\rm eff}$ of the lepton vertex, or $\kappa_e \sin^2\theta_W$, which is commonly denoted as $\sin^2\theta_{\rm eff}^{\rm lept}$. At the $Z$ pole, $\kappa_e$ is independent of the quark type. For comparisons with other measurements, the value of $\sin^2\theta_{\rm eff}^{\rm lept}$ at the $Z$ pole ${\rm Re}\,\kappa_e(s_Z)\sin^2\theta_W$ ($s_Z = M_Z^2$), is used.

Only the photon self-energy correction from fermion loops is used with the ZFITTER $Z$-amplitude form factors. The self-energy correction is a complex-valued form factor of the photon propagator, and its effect is often described as the running of the electromagnetic interaction coupling. The corrections from $W$-boson loops in the photon propagator and $Z$ propagators at the fermion-photon vertices have been combined with their gauge-dependent counter terms in the $Z$-amplitude form factors. With this reorganization of terms, all form factors are gauge invariant.

### B. QCD calculations

The Drell-Yan QCD calculations are improved by incorporating the ZFITTER form factors into the process amplitude. This provides an enhanced Born approximation (EBA) to the electroweak terms of the amplitude. The QED photon self-energy correction is included as part of the EBA. The photon amplitude influences the shape of $A_4$ away from the $Z$ pole via its interference with the axial-vector part of the $Z$ amplitude. The $\gamma$-$Z$ interference, whose cross section is proportional to $(s - M_Z^2)$, begins to dominate the total-interference cross section away from the $Z$ pole. As it dilutes measurements of $\sin^2\theta_{\rm eff}$, photonic corrections also need to be included.

The ZFITTER form factors, $\rho_{eq}$, $\kappa_e$, and $\kappa_q$ are inserted into the Born $g_A^f$ and $g_V^f$ couplings for the Drell-Yan process. To accomodate the $\kappa_{eq}$ form factor, a correction term proportional to the $(\kappa_{eq} - \kappa_e\kappa_q)$ form factor is added to the Born amplitude. The photon self-energy correction is incorporated with the photon propagator in the amplitude. Complex-valued form factors are used in the amplitude. Operationally, only the electroweak-coupling factors in the QCD cross sections are affected. To be consistent with the standard LEP $Z$-boson resonant line shape, the $Z$-boson propagator is defined as in $A_q$ (Eq. (A1)). The total-decay width $\Gamma_Z$, calculated with ZFITTER is also used.

A leading-order (LO) QCD or *tree* calculation of $\bar{A}_4$ for the process, $p\bar{p} \to \gamma^*/Z \to e^+e^-$, is used as the baseline EBA calculation with ZFITTER form factors. It is used to provide a reference for the sensitivity of $\bar{A}_4$ to QCD radiation. The CT10 [11] next-to-leading-order (NLO) parton distribution functions (PDF) provide the incoming parton flux used in all QCD calculations discussed in this section except where specified otherwise. The EBA calculation using ZFITTER form-factor tables is developed for this analysis. The EBA implementation of the form factors in the tree calculation is tested against ZGRAD2, a LO QCD calculation with electroweak radiative corrections. Only expected differences are found. The details of the tests are in Appendix B.

Two NLO calculations, RESBOS [12] and the POWHEG-BOX framework [13], are modified to be EBA-based QCD calculations. For both calculations, the boson $P_{\rm T}^2$ distribution is finite as $P_{\rm T}^2$ vanishes. The RESBOS calculation combines a NLO fixed-order calculation at high boson-$P_{\rm T}$ with the Collins-Soper-Sterman resummation formalism [14] at low boson-$P_{\rm T}$, which is an all-orders summation of large terms from gluon emission. The RESBOS calculation uses CTEQ6.6 [15] NLO PDFs. The POWHEG-BOX is a fully unweighted partonic-event generator that implements Drell-Yan production of $ee$-pairs at LO and NLO. The NLO production implements a Sudakov form factor that controls the infrared diverence at low $P_{\rm T}$, and is constructed to be interfaced with parton showering to avoid double counting. The PYTHIA 6.41 [16] parton-showering algorithm is used to produce the final hadron-level event.

At tree level, the electron angular-distribution coefficient $A_4$ is a function of the $ee$-pair rapidity ($y$) and mass ($M$): $A_4(y, M)$. The mass dependence is significant, and typically represented as the forward-backward asymmetry in $\cos\vartheta$,

$$A_{\rm fb}(M) = \frac{\sigma^+(M) - \sigma^-(M)}{\sigma^+(M) + \sigma^-(M)} = \frac{3}{8}A_4(M)\,,$$

where $\sigma^+(M)$ is the total cross section for $\cos\vartheta > 0$, and $\sigma^-(M)$ is the cross section for $\cos\vartheta < 0$. Figure 1 shows the typical behavior of $A_{\rm fb}(M)$. At $M = M_Z$, the asymmetry $A_{\rm fb}$ originates purely from $Z$ bosons, and is sensitive to $\sin^2\theta_{\rm eff}$.

Beyond leading order, the angular coefficients begin to depend on the boson $P_{\rm T}$, *i.e.*, $A_4(y, M, P_{\rm T})$. The projections $A_4(y)$ and $A_4(P_{\rm T})$ for $66 < M < 116$ GeV/$c^2$ are approximately constant except at the extremes of large $|y|$ or $P_{\rm T}$. The POWHEG-BOX events are post-processed by the PYTHIA parton showering, which adds additional boson $P_{\rm T}$, i.e., higher-order QCD corrections. While the angular-distribution coefficients of the POWHEG-BOX LO events with PYTHIA parton showering and the NLO-based coefficients are similar at low $P_{\rm T}$, they can differ at large $P_{\rm T}$.

The tree and NLO calculations of the $\bar{A}_4$ coefficient for various input values of $\sin^2\theta_W$ are shown in Fig. 2. To quantify the effects of higher-order QCD corrections on $\bar{A}_4$, the ratio $R_4 = \bar{A}_4({\rm NLO})/\bar{A}_4({\rm tree})$ is used, where NLO and tree denote $\bar{A}_4$ evaluated at NLO and at the tree level, respectively. Figure 3 shows the fractional difference $1 - R_4$ for the RESBOS and POWHEG-BOX calculations with various values of $\sin^2\theta_W$. Higher-order QCD corrections do not significantly alter $\bar{A}_A$ with respect to its value from tree-level amplitudes.

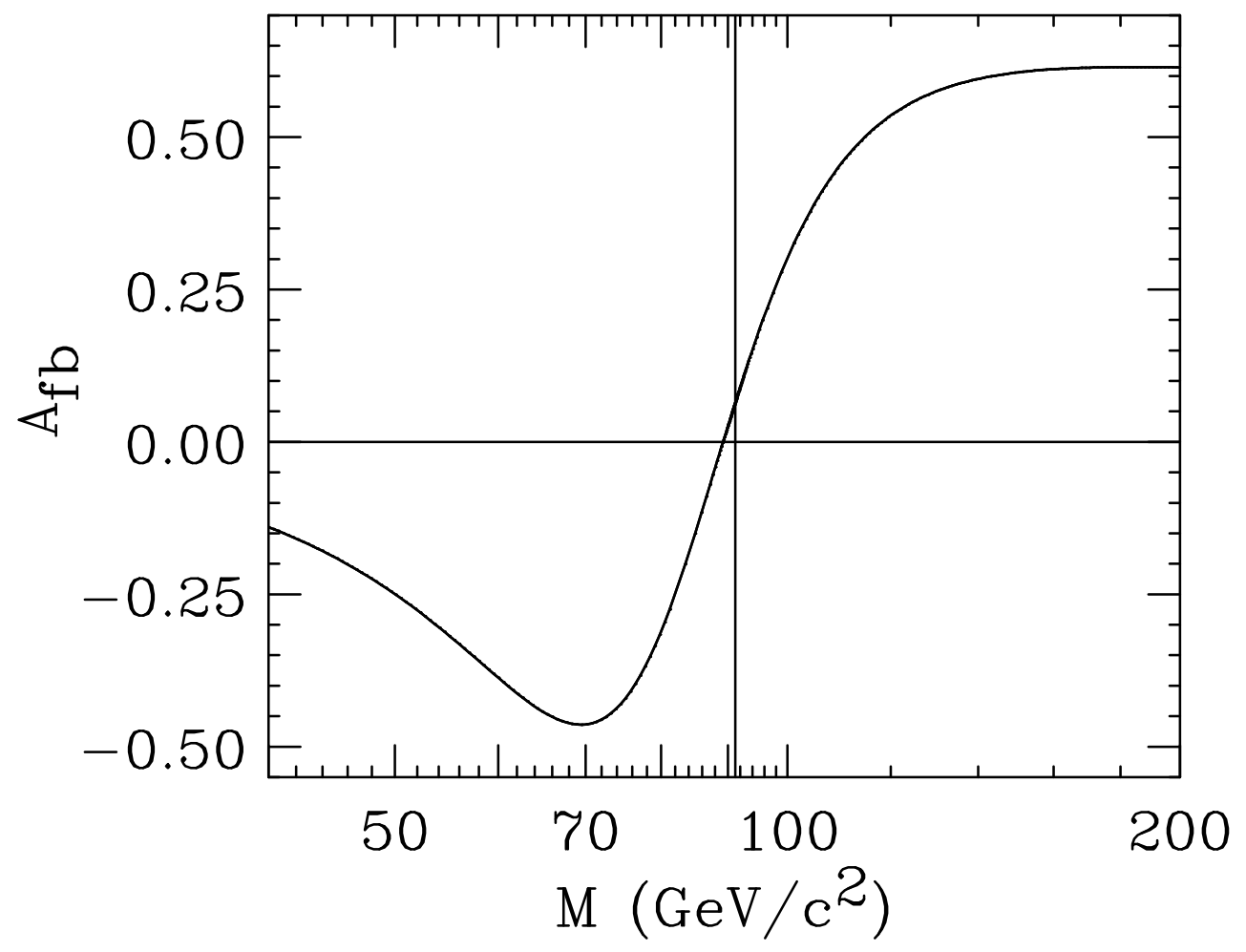

FIG. 1. Value of $A_{\text{fb}}$ as a function of mass as resulting from a tree-level calculation with $\sin^2\theta_W = 0.223$. The horizontal line corresponds to $A_{\text{fb}} = 0$ and the vertical line corresponds to $M = M_Z$.

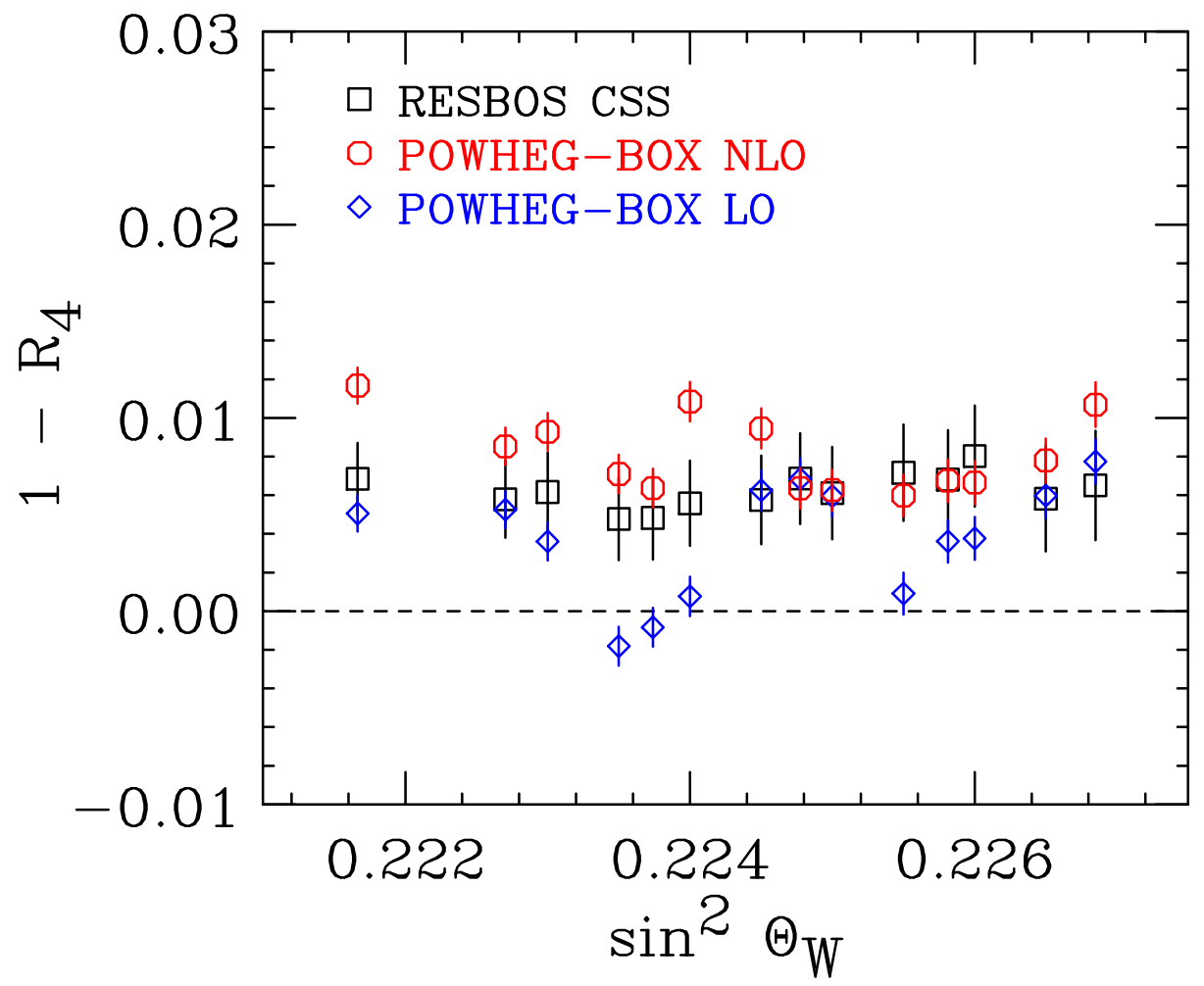

FIG. 3. $1 - R_4$ as a function of $\sin^2\theta_W$. The open squares, circles, and diamonds correspond to the RESBOS, POWHEG-BOX NLO, and POWHEG-BOX LO calculations, respectively. The POWHEG-BOX LO prediction includes higher-order QCD corrections from the parton-showering algorithm of PYTHIA.

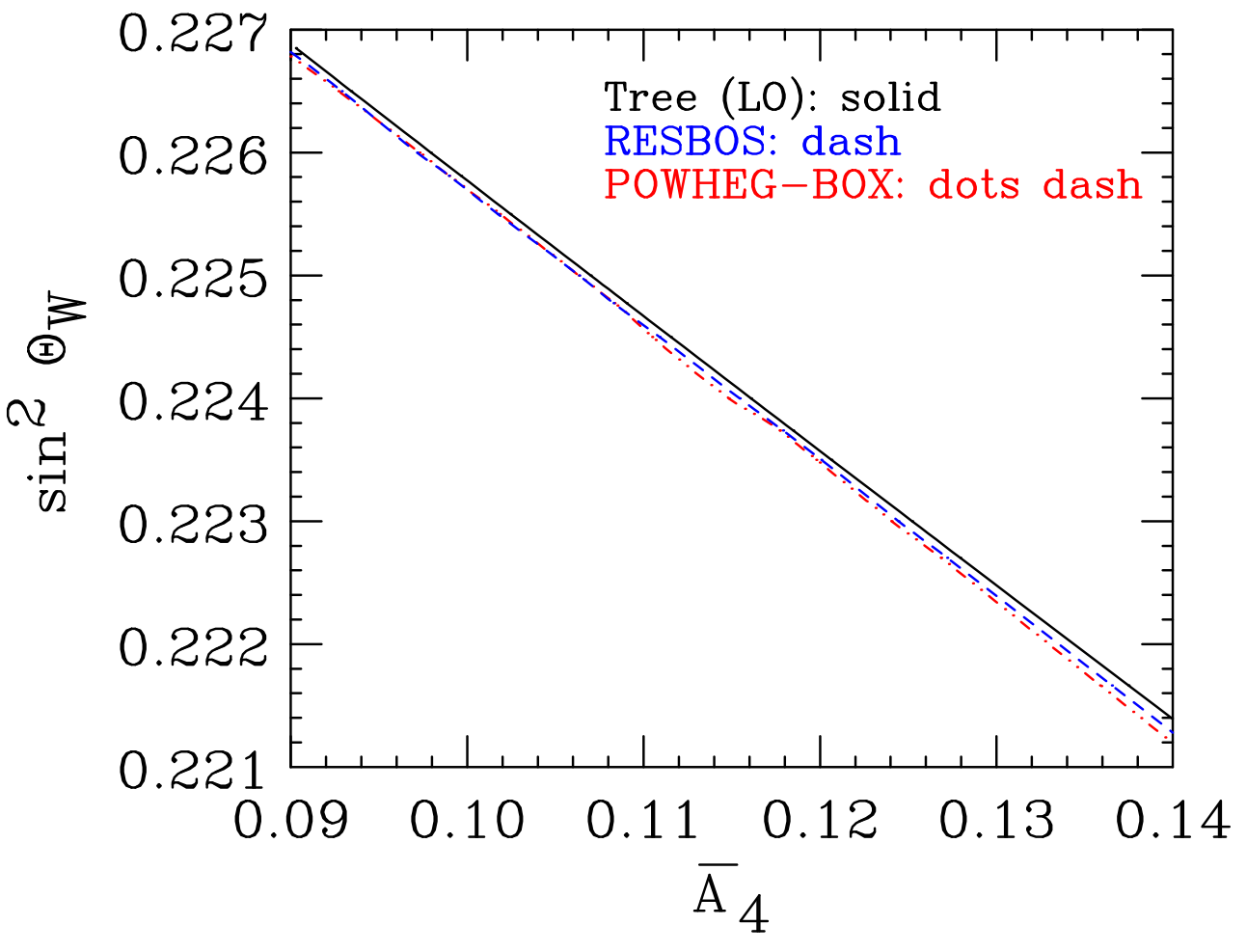

FIG. 2. Dependence of $\sin^2\theta_W$ on $\bar{A}_4$ for various $\sin^2\theta_W$ values from different QCD calculations. The tree calculation is represented by the solid (black) curve, the RESBOS calculation is represented by the dashed (blue) curve, and the POWHEG-BOX NLO calculation is represented by the dots-dashed (red) curve.

TABLE I. Measured angular coefficients [6]. The first contribution to the uncertainty is statistical, and the second systematic. The lepton-pair mass range is restricted to 66–116 GeV/$c^2$, and the mean lepton-pair $P_{\text{T}}$ values of the events in the five bins are 4.8, 14.1, 26.0, 42.9, and 73.7 GeV/$c$, respectively.

| $P_{\text{T}}$ bin (GeV/$c$) | $A_0$ ($\times 10^{-1}$) | $A_2$ ($\times 10^{-1}$) |
|---|---|---|
| 0–10 | $0.17 \pm 0.14 \pm 0.07$ | $0.16 \pm 0.26 \pm 0.06$ |
| 10–20 | $0.42 \pm 0.25 \pm 0.07$ | $-0.01 \pm 0.35 \pm 0.16$ |
| 20–35 | $0.86 \pm 0.39 \pm 0.08$ | $0.52 \pm 0.51 \pm 0.29$ |
| 35–55 | $3.11 \pm 0.59 \pm 0.10$ | $2.88 \pm 0.84 \pm 0.19$ |
| $> 55$ | $4.97 \pm 0.61 \pm 0.10$ | $4.83 \pm 1.24 \pm 0.02$ |

| $P_{\text{T}}$ bin (GeV/$c$) | $A_3$ ($\times 10^{-1}$) | $A_4$ ($\times 10^{-1}$) |
|---|---|---|
| 0–10 | $-0.04 \pm 0.12 \pm 0.01$ | $1.10 \pm 0.10 \pm 0.01$ |
| 10–20 | $0.18 \pm 0.16 \pm 0.01$ | $1.01 \pm 0.17 \pm 0.01$ |
| 20–35 | $0.14 \pm 0.24 \pm 0.01$ | $1.56 \pm 0.26 \pm 0.01$ |
| 35–55 | $-0.19 \pm 0.41 \pm 0.04$ | $0.52 \pm 0.42 \pm 0.03$ |
| $> 55$ | $-0.47 \pm 0.56 \pm 0.02$ | $0.85 \pm 0.50 \pm 0.05$ |

The RESBOS and POWHEG-BOX NLO calculations are similar and consistent. The RESBOS calculation is chosen as the default EBA-based QCD calculation of $\bar{A}_4$ with various input values of $\sin^2\theta_W$. As the POWHEG-BOX NLO program has a diverse and useful set of calculation options, it is used to estimate QCD systematic uncertainties.

## IV. EXPERIMENTAL INPUT TO $\bar{A}_4$

The value of the $\bar{A}_4$ angular-distribution coefficient is derived from the previous measurement of electron angular-distribution coefficients [6]. Elements of the measurement are summarized in this section for completeness and supplemental documentation.

The coefficients $A_0$, $A_2$, $A_3$, and $A_4$ are measured in the CS rest frame and in independent ranges of the dielectron-pair $P_{\text{T}}$. These measurements are reproduced in Table I, and are derived from a $p\bar{p}$ collision sample corresponding to an integrated luminosity is 2.1 fb$^{-1}$. The data and simulation are understood, and the modeling of the data in the simulation is accurate. The measure-ment of the angular coefficients is data driven, and fully corrected for acceptance and detector resolution.

The description of the data simulation, Sec. IV A, is presented before the description of the event sample, Sec. IV B, to aid in the discussion of the data-driven corrections to the simulation. Section IV C describes the method used to measure the angular coefficients, $A_0$, $A_2$, $A_3$, and $A_4$ in independent ranges of the dielectron-pair $P_{\mathrm{T}}$. Finally, Sec. IV D describes the method used to average the previous independent measurements of $A_4$, and to estimate the uncertainties on the combination.

### A. Data simulation

Drell-Yan pair production is simulated using the Monte Carlo event generator, PYTHIA [17], and CDF II detector-simulation programs. This simulation is only used for the measurement of the angular coefficients. PYTHIA generates the hard, leading-order QCD interaction, $q + \bar{q} \to \gamma^*/Z$, simulates initial-state QCD radiation via its parton-shower algorithms, and generates the decay $\gamma^*/Z \to l^+l^-$. The CTEQ5L [18] nucleon parton-distribution functions are used in the QCD calculations. The underlying event and boson $P_{\mathrm{T}}$ parameters are from PYTHIA tune AW (i.e., PYTUNE 101, which is a tuning to previous CDF data) [17, 19, 20]. In addition, PHOTOS 2.0 [21, 22], adds final-state QED radiation to decay vertices with charged particles (e.g. $\gamma^*/Z \to ee$). The parton-shower simulation of PYTHIA uses a QCD resummation calculation. The resulting physics model is adequate to allow data-driven adjustments to the underlying angular-distribution coefficients and other physics distributions.

The measurement of the electron angular coefficient depends on the correct modeling of the physics and both the detector acceptance and efficiency. All data efficiencies, global and particle-trajectory dependent, as well as time-dependent, are measured in the data and incorporated into the simulation. The simulation also uses the calorimeter energy scales and resolutions measured in the data. The data-driven approach is iterative with simultaneous tuning of both the generator physics-model distributions and the detector-modeling parameters that make the distributions of reconstructed quantities of simulated events match the data precisely. The tuning of the generator physics-model distributions include adjustments to both the boson production kinematics ($y$, $M$, and $P_{\mathrm{T}}$), and the lepton angular distributions ($A_0$, $A_2$, $A_3$, and $A_4$).

The PHOTOS program generates multiple photons at the $\gamma^*/Z \to ee$ vertex via a form factor to the production cross section. Soft and collinear photons are simulated to $\alpha_{em}^2$ leading-logarithmic accuracy, where $\alpha_{em}$ is the fine-structure constant. The simulation of hard, non-collinear photon emission is a full $\alpha_{em}$ matrix-element algorithm, except that the interference terms are removed to make the algorithm process-independent [22]. For the $\gamma^*/Z \to ee$ process, the interference terms are restored in an approximate way. The real and virtual photon-emission cross-section infrared divergences at each order are regularized and analytically combined to cancel the divergences. Photons with energies smaller than the default regularization energy are not generated.

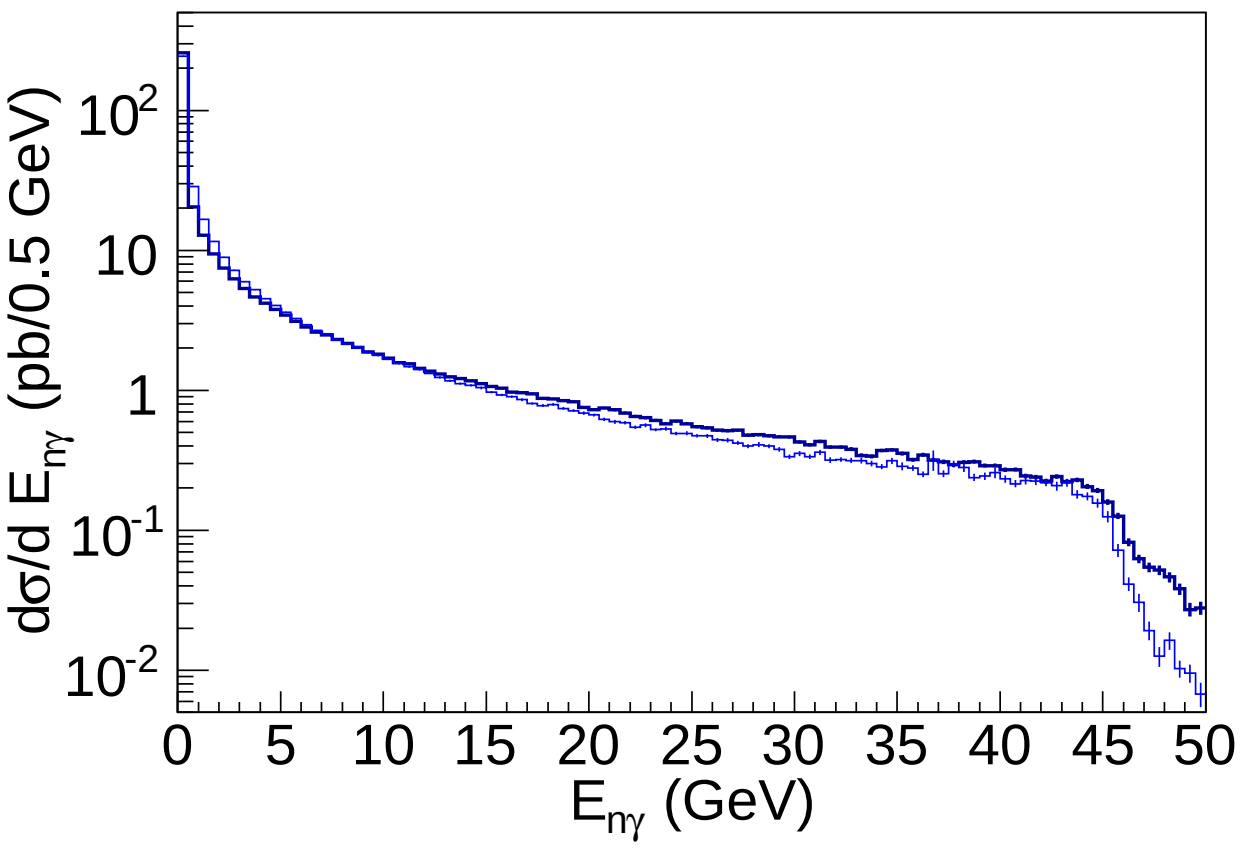

FIG. 4. Photon (cluster) energy: $E_{n\gamma}$. Events without photons are included in the lowest energy bin. The bold histogram is PYTHIA+PHOTOS. The lighter histogram is ZGRAD2. The integral of the PYTHIA+PHOTOS distribution is normalized to the ZGRAD2 total cross section.

In addition to QCD initial-state radiation, PYTHIA adds initial- and final-state QED radiation via its parton-showering algorithm. The regularization-energy threshold is very low, and most of the photons are very soft. This threshold is lower than the one in PHOTOS, so the soft-photon emission of PYTHIA is complementary to the hard-photon emission of PHOTOS.

The default implementation of PYTHIA plus PHOTOS (PYTHIA+PHOTOS) QED radiation in the CDF data-simulation infrastructure is validated with ZGRAD2 [23], a leading-order QCD Drell-Yan calculation with a $\mathcal{O}(\alpha_{em})$ matrix-element calculation for the emission of zero or one real photon. Both initial-state and final-state radiation are included. As ZGRAD2 has soft and collinear photon-regularization regions for the cancellation of divergences, these regions are excluded from comparisons with PYTHIA+PHOTOS.

The $e^+e^- +$ n$\gamma$ systems are first boosted to their center-of-momentum frames to minimize distortions to the electron and photon kinematic distributions from QCD (QED) initial-state radiation. To simplify the comparison of the multi-photon system of PYTHIA+PHOTOS to the single photon of ZGRAD2, the multi-photon system is clustered by adding up the photon momentum vectors. Events with cluster energies under 0.5 GeV, the ZGRAD2 regularization energy, are classified as events without photons. The photon (cluster) energy distributions are shown in Fig. 4. For events with photons, the smallest angle between the photon (cluster) and either lepton is denoted as $\beta$. The $\cos\beta$ distribution is shown in Fig. 5. The overall consistency is good. Differences are expected as the PYTHIA+PHOTOS correction is $\mathcal{O}(\alpha_{em}^2)$ or larger, while the ZGRAD2 correction is $\mathcal{O}(\alpha_{em})$.

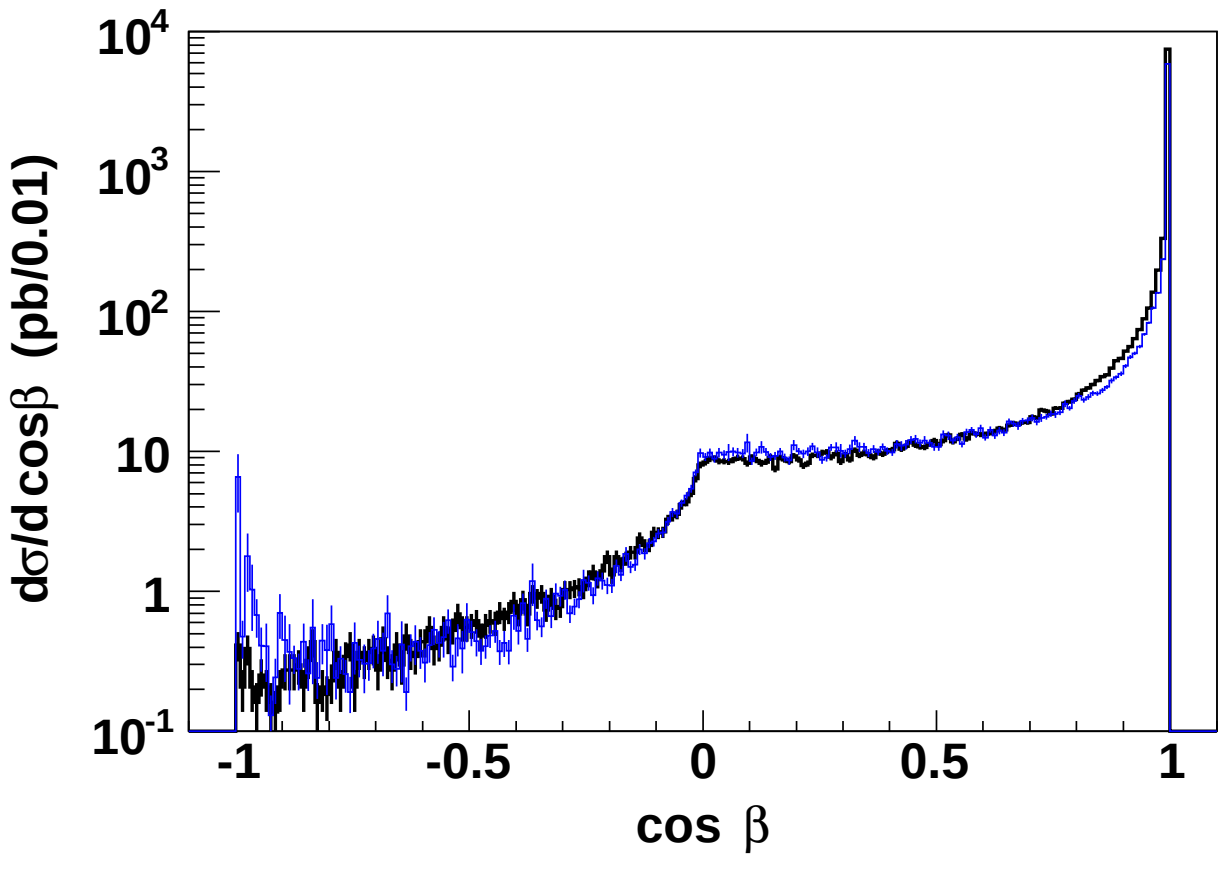

FIG. 5. Separation between the photon (cluster) and the nearest lepton: $\cos\beta$. The bold histogram is PYTHIA+PHOTOS. The lighter histogram is ZGRAD2. The integral of the PYTHIA+PHOTOS distribution within $0 < \cos\beta < 0.8$ is normalized to the corresponding ZGRAD2 cross section.

### B. Measurement event sample

The CDF experimental apparatus is a general-purpose detector [24] at the Fermilab Tevatron $p\bar{p}$ collider whose center-of-momentum (cm) energy is 1.96 TeV. The positive $z$-axis is directed along the proton direction. For particle trajectories, the polar angle $\theta_{\rm cm}$ is relative to the proton direction and the azimuthal angle $\phi_{\rm cm}$ is oriented about the beamline axis with $\pi/2$ being vertically upwards. The component of the particle energy transverse to the beamline is defined as $E_{\rm T} = E\sin\theta_{\rm cm}$. The pseudorapidity of a particle trajectory is $\eta = -\ln\tan(\theta_{\rm cm}/2)$. Detector coordinates are specified as $(\eta_{\rm det}, \phi_{\rm cm})$, where $\eta_{\rm det}$ is the pseudorapidity relative to the detector center ($z = 0$).

The central charged-particle tracking-detector (tracker) is a 3.1 m long, open-cell drift chamber [25] that radially extends from 0.4 to 1.4 m. Between the Tevatron beam pipe and the central tracker is a 2 m long silicon vertex-tracker [26]. Both trackers are immersed in a 1.4 T axial magnetic field. Outside the central tracker is a central barrel calorimeter [27, 28] that covers the region $|\eta_{\rm det}| < 1.1$. The forward end-cap regions are covered by the end-plug ("plug") calorimeters [29–31] that cover the regions $1.1 < |\eta_{\rm det}| < 3.5$. Both the central and plug calorimeters are segmented into electromagnetic and hadronic sections. The electromagnetic sections of both calorimeters have preshower and shower-maximum detectors for electron identification. The silicon tracker, in conjunction with the plug shower-maximum detector, provides tracking coverage in the plug region to $|\eta_{\rm det}|$ of about 2.8. As $|\eta_{\rm det}|$ increases for plug-region tracks, the transverse track length within the magnetic field decreases, resulting in increasingly poorer track-curvature resolutions.

Events are required to contain two electron candidates having a pair mass in the $Z$-boson region of 66–116 GeV/$c^2$. Electrons in both the central and plug calorimeters are used. The events are classified into three di-electron topologies: CC, CP, and PP, where C (P) denotes that the electron is detected in the central (plug) calorimeter. Electrons are required to have an associated track, pass standard selection and fiducial requirements [24], and be isolated from other calorimeter activity. The electron kinematic variables are based on the electron energy measured in the calorimeters and the track direction. The kinematic and fiducial regions of acceptance for electrons in the three topologies are summarized below.

1. Central–Central (CC)
   - $E_{\rm T} > 25$ (15) GeV for electron 1 (2)
   - $0.05 < |\eta_{\rm det}| < 1.05$
2. Central–Plug (CP)
   - $E_{\rm T} > 20$ GeV for both electrons
   - Central electron: $0.05 < |\eta_{\rm det}| < 1.05$
   - Plug electron: $1.2 < |\eta_{\rm det}| < 2.8$
3. Plug–Plug (PP)
   - $E_{\rm T} > 25$ GeV for both electrons
   - $1.2 < |\eta_{\rm det}| < 2.8$

The CC-electron $E_{\rm T}$ selection is asymmetric, with electron 1 having the highest $E_{\rm T}$. The asymmetric selection, an optimization from the previous measurement of electron angular-distribution coefficients, improves the acceptance in the electron phase space [6]. The PP-electron candidates, required to be in the same end of the CDF II detector, extend the rapidity coverage to $|y| \approx 2.9$. The kinematic limit of $|y|$ for the production of $ee$-pairs at the $Z$-boson mass is 3.1. The acceptance is limited for PP-topology Drell-Yan electrons on opposite ends of the CDF II detector; the dielectrons tend to be at low $ee$-pair rapidities, and are overwhelmed by the QCD di-jet backgrounds.

The numbers of events passing all requirements in the CC, CP, and PP topologies are 51 951, 63 752, and 22 469, respectively. All requirements in the association of charged-particle tracks to both final-state electrons significantly reduces the backgrounds coming from QCD, the electroweak (EWK) processes of $WW$, $WZ$, $ZZ$, $t\bar{t}$, $W$+jets, and also $Z \to \tau^+\tau^-$. The QCD background is primarily from dijets where a particle in a jet is misidentified as an electron or is an electron from a photon conversion. The high-$E_{\rm T}$ electron sources have at least one real electron. The second electron is either a real second electron or a fake one. The backgrounds and the methods used to determine them are described further in previous measurements [6, 32]. The QCD backgrounds, determined from the same dielectron sample used for the measurement, constitute 0.3% of the sample. The EWK backgrounds are derived from PYTHIA [17] samples with detector simulation, and amount to 0.2%. The fraction of QCD plus EWK backgrounds is approximately constant over $\cos\vartheta$ for each topology. Background-subtracted distributions are used in measurements.

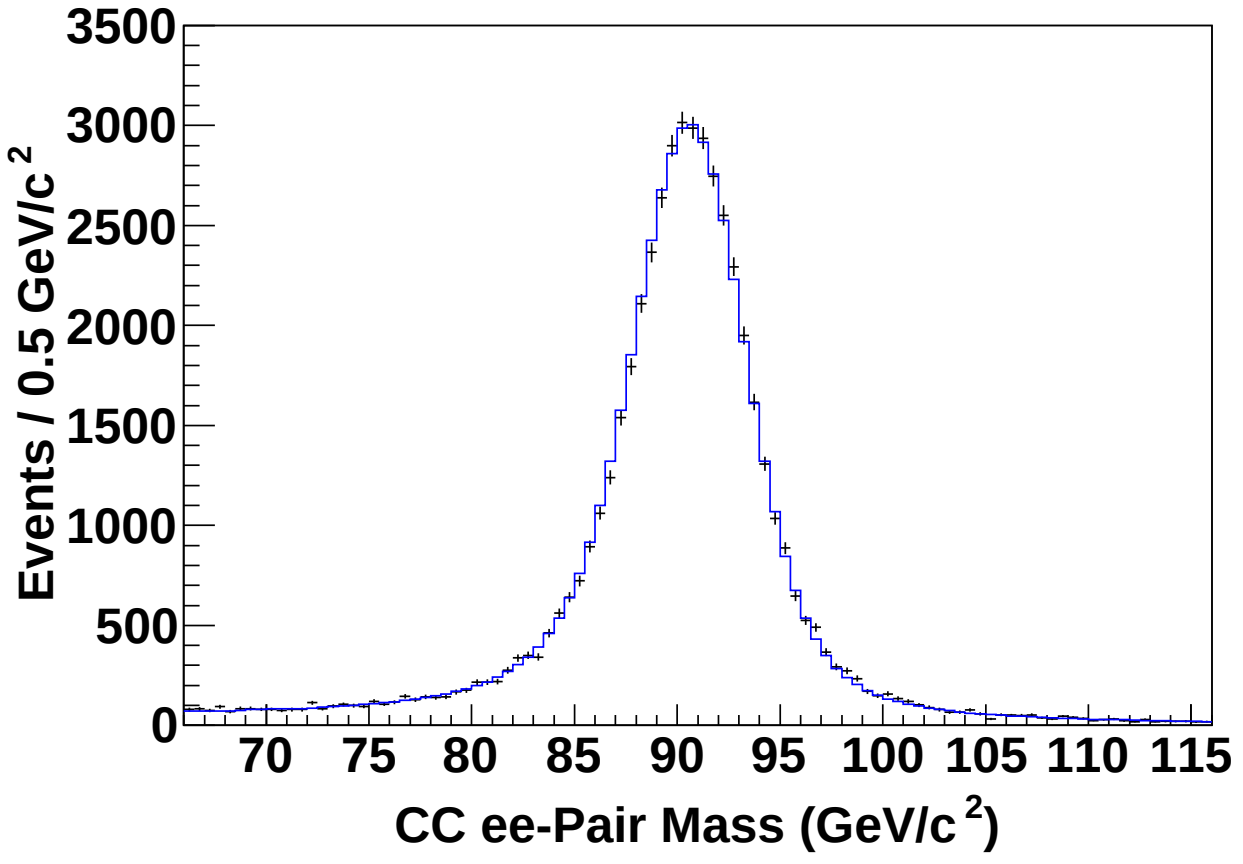

FIG. 6. The CC-topology $ee$-pair mass distribution. The crosses are the data and the histogram is the simulation.

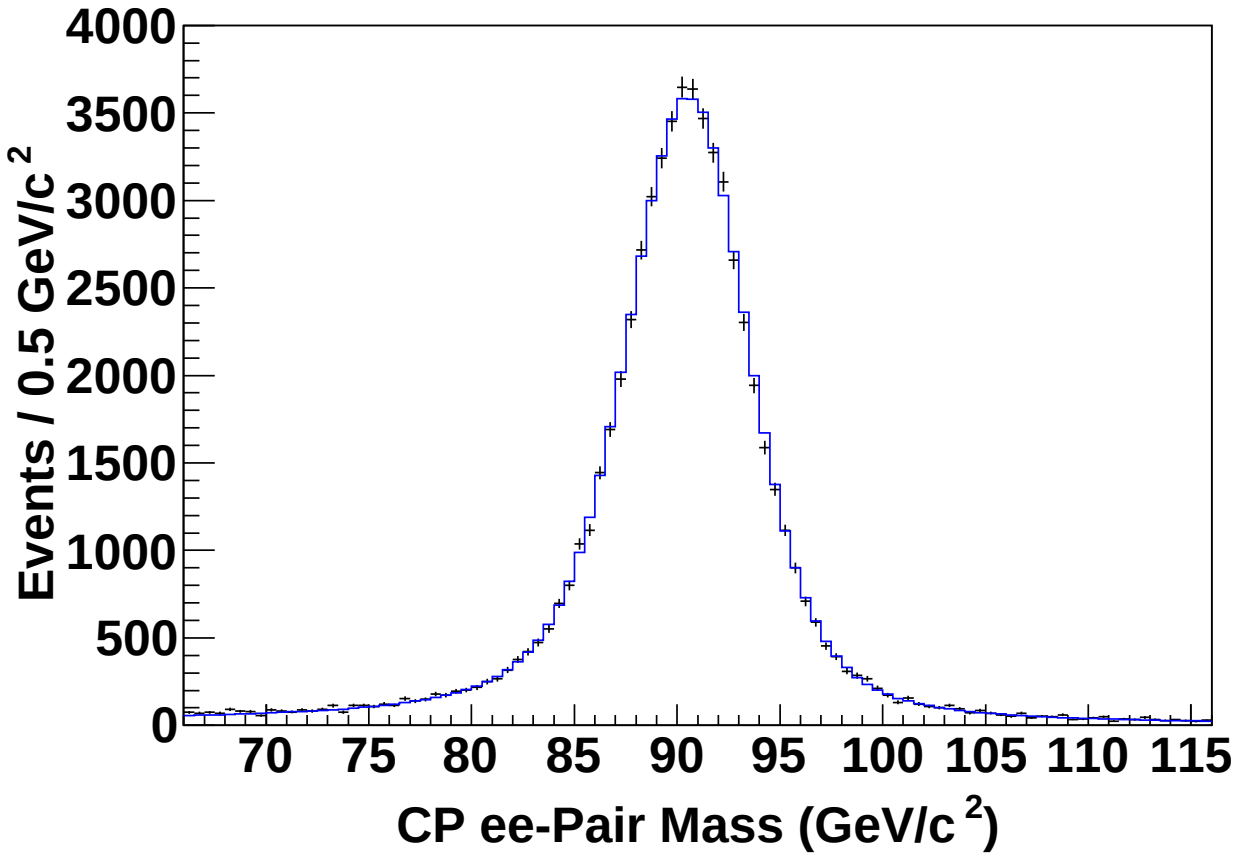

FIG. 7. The CP-topology $ee$-pair mass distribution. The crosses are the data and the histogram is the simulation.

The online-event selection and electron-identification efficiencies are measured as functions of $\eta_{\rm det}$ for both central and plug electrons. The measured efficiencies are incorporated in the simulation as scale factors (event weights). Plug-electron efficiencies are separately measured for the CP and PP electrons. A significant fraction of the PP-toplology electrons are in more forward regions of the calorimeter relative to those of the CP topology. The efficiencies for electrons to be identified in the plug calorimeter particularly in the very forward regions, have a significant time-dependence (due to increasing instantaneous luminosities) which are measured and incorporated into the simulation.

Corrections to the simulated-event electron energy-scales and resolutions are determined using both the $ee$-pair mass and electron-$E_{\rm T}$ distributions. The energy scales and resolutions of the simulation are adjusted so that both the simulated-electron $E_{\rm T}$ distributions and the $ee$-pair mass distributions are matched to the observed distributions [32]. The central- and plug-electron energy scales are accurately constrained by the three independent $ee$-pair topologies. Figures 6, 7, and 8 show the $ee$-pair mass distributions for the CC, CP, and PP topologies, respectively. The simulated-data to data $\chi^2$ for the CC-, CP-, and PP-topology $ee$-pairs are 117, 126, and 127, respectively, for 100 bins. The event count of the simulated data is normalized to that of the data, and only statistical uncertainties are used in the calculation.

The Collins-Soper frame angle, $\cos\vartheta$ [7], is reconstructed using these laboratory-frame quantities: the lep-ton energies ($E$), the lepton momenta along the beam line ($P_z$), the dilepton mass ($M$), and the dilepton transverse momentum ($P_{\rm T}$). The angle of the negatively-charged lepton is

$$\cos\vartheta = \frac{l_+^- l_-^+ - l_-^- l_+^+}{M\sqrt{M^2 + P_{\rm T}^2}} ,$$

where $l_\pm = (E \pm P_z)$ and the $+$ $(-)$ superscript specifies that $l_\pm$ is for the positively- (negatively-)charged lepton. A similar expression is used for $\varphi$. For plug electrons, charge identification is not used because of significant charge misidentification probability at large $|\eta_{\rm det}|$. As an interchange of the $e^-$ with the $e^+$ changes the sign of $\cos\vartheta$, $|\cos\vartheta|$ is used for the PP-topology dielectrons. For CP-topology dielectrons, the central-electron charge determines whether the $e^-$ is the central or plug electron. For the CC- and CP-topology dielectrons, the charge-misidentification probabilities are 0.3% and 0.4% respectively.

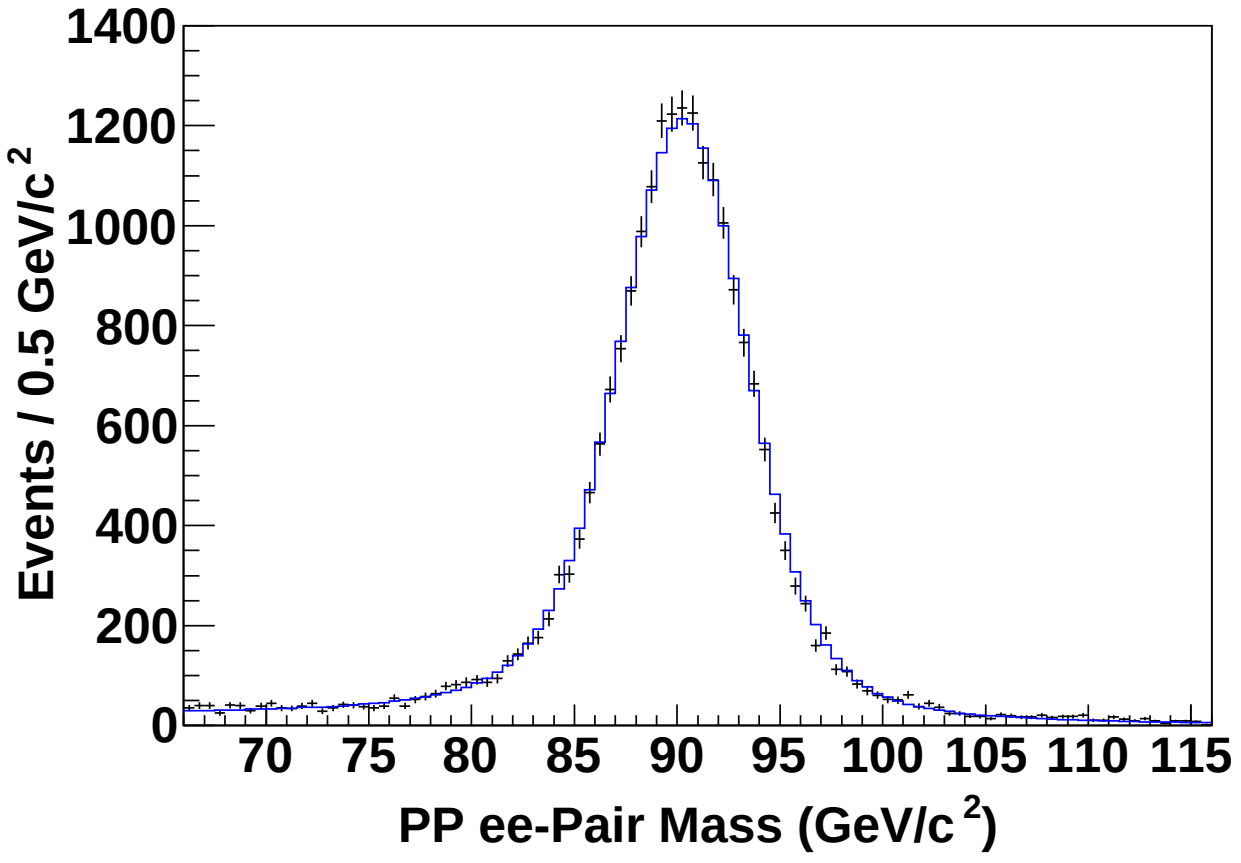

FIG. 8. The PP-topology $ee$-pair mass distribution. The crosses are the data and the histogram is the simulation.

The $\cos\vartheta$ bias and resolution of the observed events are estimated using the simulation. The bias $\Delta\cos\vartheta$, is the difference between the true $\cos\vartheta$ before final-state QED radiation and the measurement. The $\Delta\cos\vartheta$ distribution is affected by the electron-energy resolution of the calorimeters and electron-charge misidentification. The effect of calorimeter energy-resolution smearing is small for all dielectron topologies. The bias distribution has a narrow non-Gaussian central core centered at zero with less than 1% rms deviation. The calorimeters have a negligible effect on the mean of the bias but dominate the resolution. Charge misidentification in the CC- and CP-dielectron topologies contributes a relatively flat background with a negligible bias.

### C. Angular coefficient measurement

The angular distribution integrated over $\varphi$ is

$$N(\vartheta, A_0, A_4) \propto 1 + \cos^2\vartheta + A_0\,\frac{1}{2}(1 - 3\cos^2\vartheta) + A_4\cos\vartheta. \quad (2)$$

In each $P_{\rm T}$ bin, this distribution is modified by the acceptance and resolution of the detector into the observed $\cos\vartheta$ distribution. The simulated events used to model the $\cos\vartheta$ distribution are selected as data. The underlying $A_0$ and $A_4$ values in the simulation physics model are simultaneously varied until the simulated $\cos\vartheta$ distributions match the corresponding data distributions. The variation is accomplished with an event weight

$$w = \frac{N(\vartheta, A_0', A_4')}{N(\vartheta, A_0, A_4)}\,.$$

The base physics-model angular coefficients are denoted as $A_0$ and $A_4$, and variations to them are denoted as $A_0'$ and $A_4'$. The best-fit values for $A_0'$ and $A_4'$ are determined using a binned log-likelihood fit between the data and simulation. The event normalization of the simulation sample relative to the data is a parameter in the log-likelihood fit as the detector acceptance depends on $A_0$ and $A_4$. The log-likelihood of each dielectron topology is separately evaluated and then combined into a joint probability-density function.

The best-fit values of $A_0'$ and $A_4'$ for each $P_{\rm T}$ bin are incorporated into the physics model prior to the determination of $\varphi$-based angular coefficients. The angular distribution integrated over $\cos\vartheta$ is

$$N(\varphi, A_2, A_3) \propto \frac{8}{3} + \frac{2}{3}A_2\,\cos 2\varphi + \frac{\pi}{2}A_3\,\cos\varphi.$$

The $A_5$ and $A_7$ terms, expected to be relatively small [8], are dropped. The best-fit values to $A_2$ and $A_3$, denoted as $A_2'$ and $A_3'$ respectively, are also obtained using the same method as for $A_0'$ and $A_4'$. The fits to the observed $\cos\vartheta$ and $\varphi$ distributions are iterated to obtain the final values of $A_0'$, $A_1'$, $A_3'$, and $A_4'$ for each $P_{\rm T}$ bin. The measurements are fully corrected for detector acceptance and resolution.

### D. $A_4$ average

The measured values of $A_0$, $A_2$, $A_3$, and $A_4$ (Table I) are incorporated into the physics model. The one-dimensional $\cos\vartheta$ distribution of events with $ee$-pair masses in the range 66–116 GeV/$c^2$ has the functional form, $N(\vartheta, \bar{A}_0, \bar{A}_4)$ (Eq. (2)). The best fit to the distribution for the functional form yields the parameters $\bar{A}_0 = 0.0514 \pm 0.0010$ and $\bar{A}_4 = 0.1100 \pm 0.0008$, where the uncertainties are due to the limited size of the simulated sample. The parameters are uncorrelated because their angular functions are orthogonal Legendre polynomials. These angular-coefficient parameters are the cross-section weighted averages based on the measurements. Without the data-driven corrections, the baseline (PYTHIA) model gives $\bar{A}_4 = 0.1128$ and $\bar{A}_0 = 0.0304$.

Experimental uncertainties of $\bar{A}_4$ are evaluated directly from the observed and simulated $\cos\vartheta$ distributions of events selected for the angular-distribution measurement. As the simulated distributions include the measured values of $A_0$, $A_2$, $A_3$, and $A_4$, the variations of $A_0'$ and $A_4'$ are via scale factors to $\bar{A}_0$ and $\bar{A}_4$ of the simulation physics model. The best-fit values from the log-likelihood fits are $\bar{A}_0 = 0.0497 \pm 0.0073$ and $\bar{A}_4 = 0.1095 \pm 0.0079$, and the central values are consistent with the cross-section weighted averages. The uncertainties reflect the limited data-sample size. The correlation coefficient between the uncertainty of $\bar{A}_4$ and $\bar{A}_0$ or the simulation normalization is under 0.01. The angular function of the $A_4$ coefficient is an odd function in $\cos\vartheta$, and it is also orthogonal to $1 + \cos^2\vartheta$.

The experimental value of $\bar{A}_4$ used to infer $\sin^2\theta_W$ is

$$\bar{A}_4 = 0.1100 \pm 0.0079,$$

where the central value is the cross-section weighted average, and its uncertainty is the statistical uncertainty from the log-likelihood fit.

The $\cos\vartheta$ distribution for the combined CC- and CP-topology dielectrons is shown in Fig. 9. The comparison of the simulation with the data yields a $\chi^2$ of 44.8 for 50 bins. The yield of simulated events is determined by the fit. For the separate CC- and CP-topology $\cos\vartheta$ distributions, the comparison between the simulation and the data yields a CC-topology $\chi^2$ of 49.0 for 50 bins, and a CP-topology $\chi^2$ of 46.9 for 46 bins. Figure 10 shows the $\cos\vartheta$ distribution of the PP topology. The comparison of the simulation with the data yields a $\chi^2$ of 31.7 for 35 bins. The CC and CP topologies are the ones that mainly constrain the fit for $\bar{A}_0$ and $\bar{A}_4$. The PP topology helps to constrain the simulation event normalization.

The observed $\varphi$ distributions are also well described by the simulation. Figure 11 shows the distribution for the combined CC and CP $ee$-pair topologies. The comparison of the simulation with the data yields a $\chi^2$ of 51.5 for 50 bins. For the separate CC- and CP-topology $\varphi$ distributions, the $\chi^2$ between the simulation and the data are 56.1 and 46.9, respectively, for 50 bins. Figure 12 shows the $\varphi$ distribution for events in the PP topology.

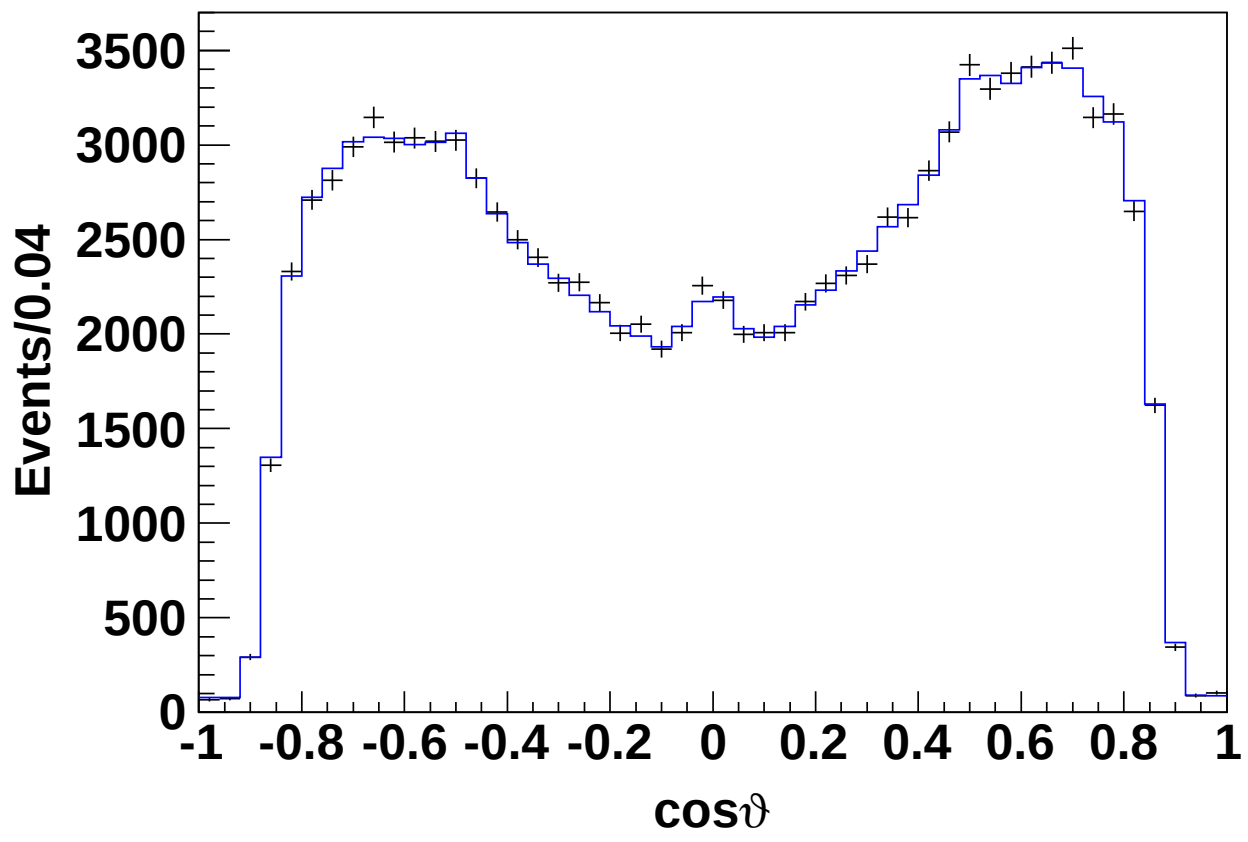

FIG. 9. The observed $\cos\vartheta$ distribution for the combined CC and CP topologies. The crosses are the background-subtracted data, and the solid histogram is the simulation.

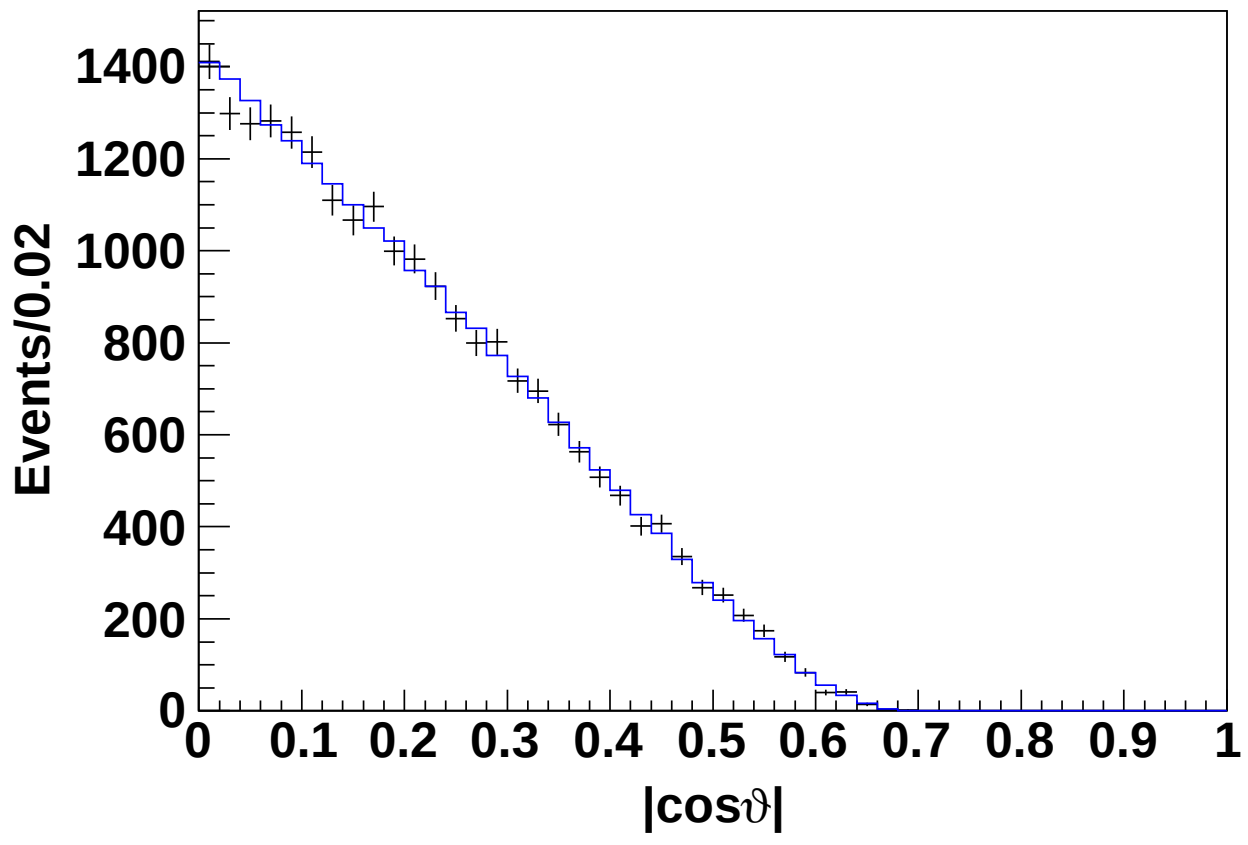

FIG. 10. The observed $|\cos\vartheta|$ distribution for the PP topology. The crosses are the background-subtracted data, and the solid histogram is the simulation.

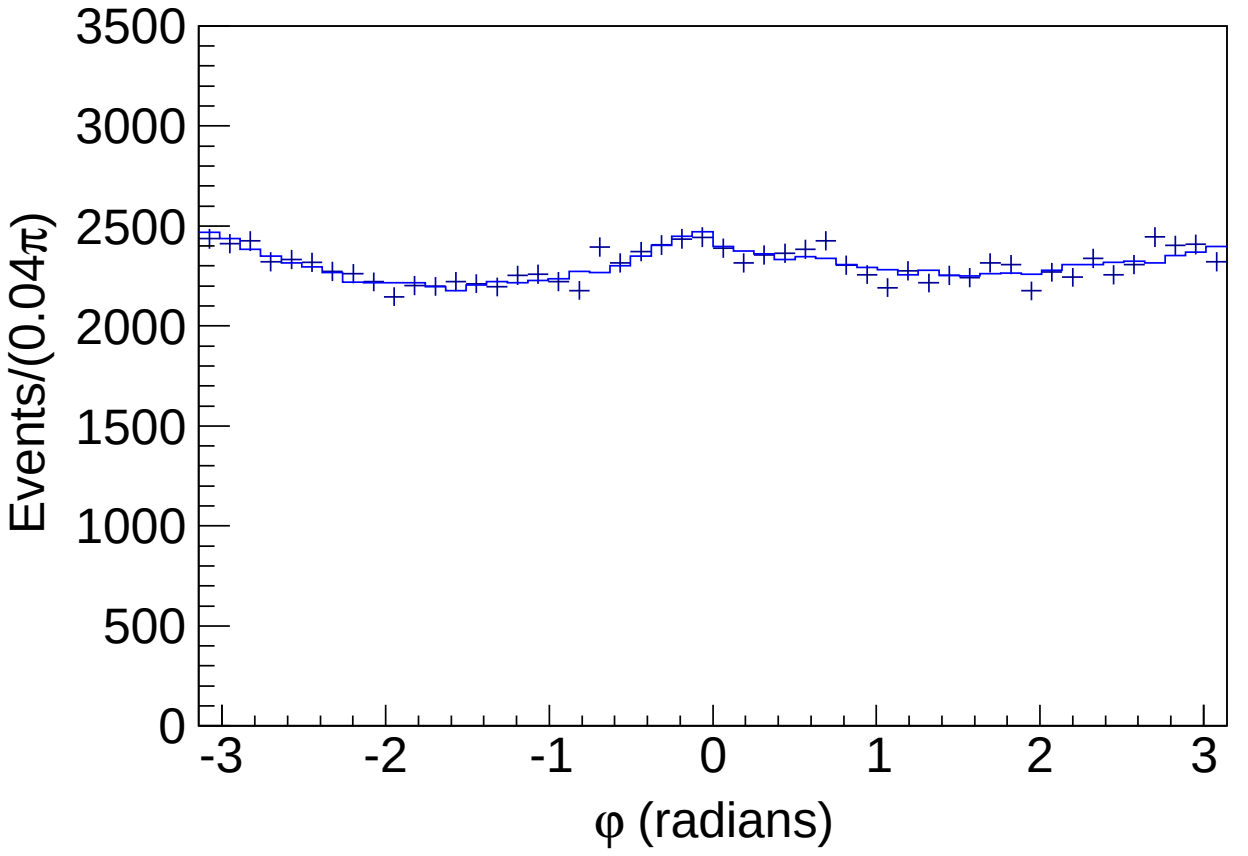

FIG. 11. The observed $\varphi$ distribution for the combined CC and CP topologies. The crosses are the background-subtracted data, and the solid histogram is the simulation.

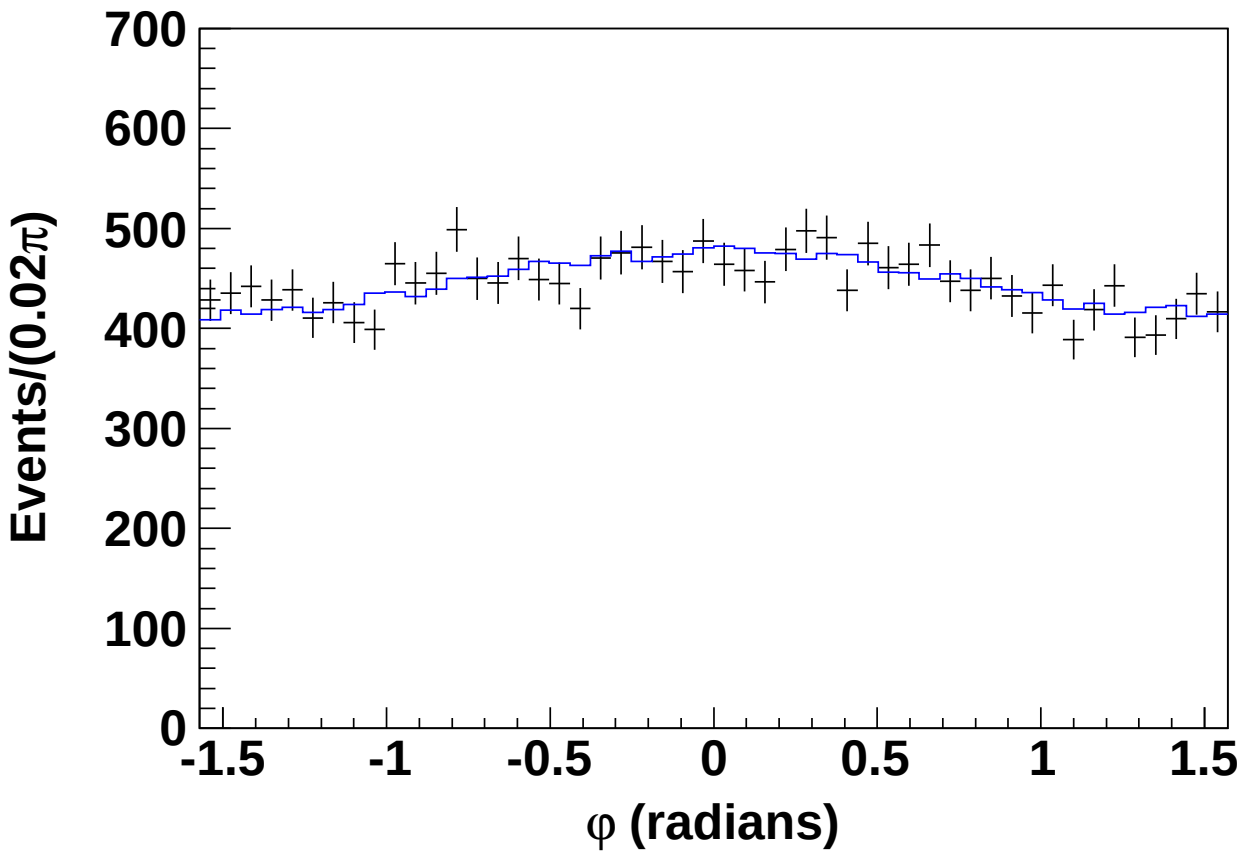

FIG. 12. The observed $\varphi$ distribution of electrons for the PP topology. The crosses are the background-subtracted data, and the solid histogram is the simulation.

The comparison of simulation with the data yields a $\chi^2$ of 47.4 for 50 bins.

## V. SYSTEMATIC UNCERTAINTIES

The systematic uncertainties on the inference of $\sin^2\theta_W$ (or $M_W$) contain contributions from both the experimental input for $\bar{A}_4$ and the predictions of $\bar{A}_4$ for various input values of $\sin^2\theta_W$. The prediction uncertainties dominate. Both the experimental and prediction systematic uncertainties are small compared to the experimental statistical uncertainty.

### A. Experimental input

The $\bar{A}_4$ angular-coefficient uncertainties considered include the simulation energy scale, the background estimates, and the single-electron selection and tracking-efficiency measurements.

The central- and plug-electron energy scales for the simulation are accurately constrained by the data. Their residual uncertainties correspond to an estimated uncertainty for the $\bar{A}_4$ coefficient of $\pm 0.0003$. This is not completely independent of the experimental statistical uncertainty, but is included in quadrature with the other experimental systematic uncertainties.

The largest independent uncertainty is from the background subtraction. It is estimated by varying the fraction of the default background that is subtracted, then re-fitting the observed $\cos\vartheta$ distribution for a modified best-fit value of $\bar{A}_4$. The level of background subtracted from the data is varied so that the change in the corresponding likelihood value corresponds to the nominal one-standard-deviation change of the results with respect to the central value. The result is $\Delta\bar{A}_4 = \pm 0.0003$.

The measured single-electron efficiences incorporated in the simulation have uncertainties. When propagated to the $\cos\vartheta$ bins, the fractional uncertainties of the CC, CP, and PP topologies are relatively constant. The levels of uncertainty for the CC, CP, and PP topology yields are 0.9%, 0.6%, and 4%, respectively. The PP-topology electron acceptance extends into very forward regions of the plug calorimeter, and signficantly beyond that for CP-topology electrons. As measurements are difficult in this far forward region, the PP uncertainty is larger. Since the same single-electron measurements are used in each bin, they are treated as 100% correlated across the $\cos\vartheta$ bins. To estimate uncertainties, the overall dielectron-topology efficiency is rescaled within its uncertainty prior to log-likelihood fits of the observed $\cos\vartheta$ distribution. This is equivalent to a systematic offset in its event normalization relative to the other topologies. The uncertainty on the $\bar{A}_4$ coefficient from this source is found to be negligible. Because the angular function of the $\bar{A}_4$ coefficient $\cos\vartheta$, is odd, the normalization of the simulated events and $\bar{A}_4$ are nearly uncorrelated in all fits.

### B. Predictions

The QCD mass-factorization and renormalization scales and uncertainties in the CT10 PDFs affect the calculated value of $\bar{A}_4$. The corresponding systematic uncertainties on $\bar{A}_4$ are evaluated using POWHEG-BOX NLO. As the RESBOS calculation is chosen as the default for $\bar{A}_4$, the associated uncertainty is also included in the overall systematic uncertainty.

In all QCD calculations, the mass-factorization and renormalization scales are both set to the *ee*-pair mass. To evaluate the effect on $\bar{A}_4$ from different scales, the running scales are varied independently by a factor ranging from 0.5 to 2 in the calculations. The largest observed deviation in $\bar{A}_4$ from the default value is the QCD-scale uncertainty. This uncertainty is $\Delta\bar{A}_4(\text{QCD scale}) = \pm 0.0004$.

The CT10 set of 26 eigenvector pairs of 90% C.L. uncertainty PDFs are used to evaluate the effect of PDF uncertainties on $\bar{A}_4$: The quadrature sum of the PDF uncertainties to $\bar{A}_4$ from each pair gives the total PDF uncertainty. The 68% C.L. uncertainty to $\bar{A}_4$ is obtained by rescaling the 90% C.L. uncertainty down by a factor of 1.645 to give $\Delta\bar{A}_4(\text{PDF}) = \pm 0.0026$.

The default RESBOS calculation of the $\bar{A}_4$ coefficent for various input values of $\sin^2\theta_W$ yields coefficent values 0.5–0.8% larger than the baseline tree calculation. The POWHEG-BOX calculations are slightly different. A conservative systematic uncertainty of ±1% is assigned for differences, and this is denoted as the EBA uncertainty.

In summary, the total systematic uncertainty from the QCD mass-factorization and renormalization scales, and uncertainties in the CT10 PDFs is $\Delta\bar{A}_4(\text{QCD}) = \pm 0.0026$. The EBA uncertainty is $\Delta\bar{A}_4(\text{EBA}) = \pm 0.01\bar{A}_4$. These prediction uncertainties are combined in quadrature. At the measured value of $\bar{A}_4$ (0.1100), the total prediction uncertainty is ±0.0029.

## VI. RESULTS

The fully-corrected value of the $\bar{A}_4$ coefficient for this analysis is

$$\bar{A}_4 = 0.1100 \pm 0.0079 \pm 0.0004,$$

where the first contribution to the uncertainty is statistical and the second systematic. Prediction uncertainties are separated from experimental uncertainties, but for the total uncertainties of derived results presented in this section, all uncertainties are combined in quadrature.

The $A_4$ angular coefficient is directly sensitive to the $\sin^2\theta_{\text{eff}}$ parameter at the lepton and quark vertices of the Drell-Yan amplitude. However, it is most sensitive to the effective-mixing parameter at the lepton vertex, and consequently, the $A_4$ coefficient is primarily a measure of $\sin^2\theta^{\text{lept}}_{\text{eff}}$. The standard model (SM) provides the means to express the effective-mixing parameters in terms of its static parameters and the collision dynamics, to map the correspondence between the effective-mixing parameters and the angular coefficient $A_4$,

$$\text{SM}(\sin^2\theta_W) \overset{\text{EWK}}{\longmapsto} \sin^2\theta_{\text{eff}}(s) \overset{\text{QCD}}{\longleftrightarrow} A_4(s),$$

and to interpret measurements of this coefficient in terms of the fundamental $W$-boson mass, $M_W$, or the $\sin^2\theta_W$ parameter. The symbol EWK denotes electroweak radiative corrections, and the symbol QCD denotes EBA-based QCD calculations. For the $\bar{A}_4$ coefficient, the kinematic dependencies of the $\sin^2\theta_{\text{eff}}(s)$ parameters are averaged by the integration over the $\sqrt{s}$ range of 66–116 GeV. Over this range, the predicted differences between the effective-leptonic and effective-quark mixing parameters are under 0.0005 in magnitude. The interpretation of the measured $\bar{A}_4$ coefficient in terms of the $\sin^2\theta_W$ or $M_W$ parameter is interesting, but model dependent. Under different standard-model contexts, the same value of an effective-mixing parameter can be associated with different values of the $\sin^2\theta_W$ parameter.

The RESBOS predictions of $\bar{A}_4$ for various values of the $M_W$ (or $\sin^2\theta_W$) parameter are shown in Fig. 13 along with the observed value. The intersection of the measured value with the prediction can be interpreted as the indirect measurement of $M_W$ or $\sin^2\theta_W$ within the context of standard-model assumptions specified in Appendix A:

$$\sin^2\theta_W = 0.2246 \pm 0.0009$$
$$M_W(\text{indirect}) = 80.297 \pm 0.048 \text{ GeV}/c^2,$$

where the uncertainty includes both measurement and prediction uncertainties. The experimental statistical uncertainty for the value of $M_W$ is $\pm 0.045$ GeV/$c^2$. The systematic uncertainty, predominantly from the prediction, is $\pm 0.017$ GeV/$c^2$. The corresponding statistical and systematic uncertainties for the value of $\sin^2\theta_W$ are $\pm 0.0009$ and $\pm 0.0003$, respectively. The other $W$-mass measurements shown in Fig. 13 are from combinations of the Tevatron, and the LEP-1 and SLD measurements [2]:

$$\begin{aligned} M_W &= 80.385 \pm 0.015 \text{ GeV}/c^2, \quad \text{direct} \\ &= 80.365 \pm 0.020 \text{ GeV}/c^2, \quad Z \text{ pole}, \end{aligned}$$

where direct refers to the combination of LEP-2 and Tevatron $W$-mass measurements, and $Z$ pole is an indirect measurement from electroweak standard-model fits to LEP-1 and SLD $Z$-pole measurements with the top-quark mass measurement. Figure 14 shows the comparison of these $W$-boson mass results.

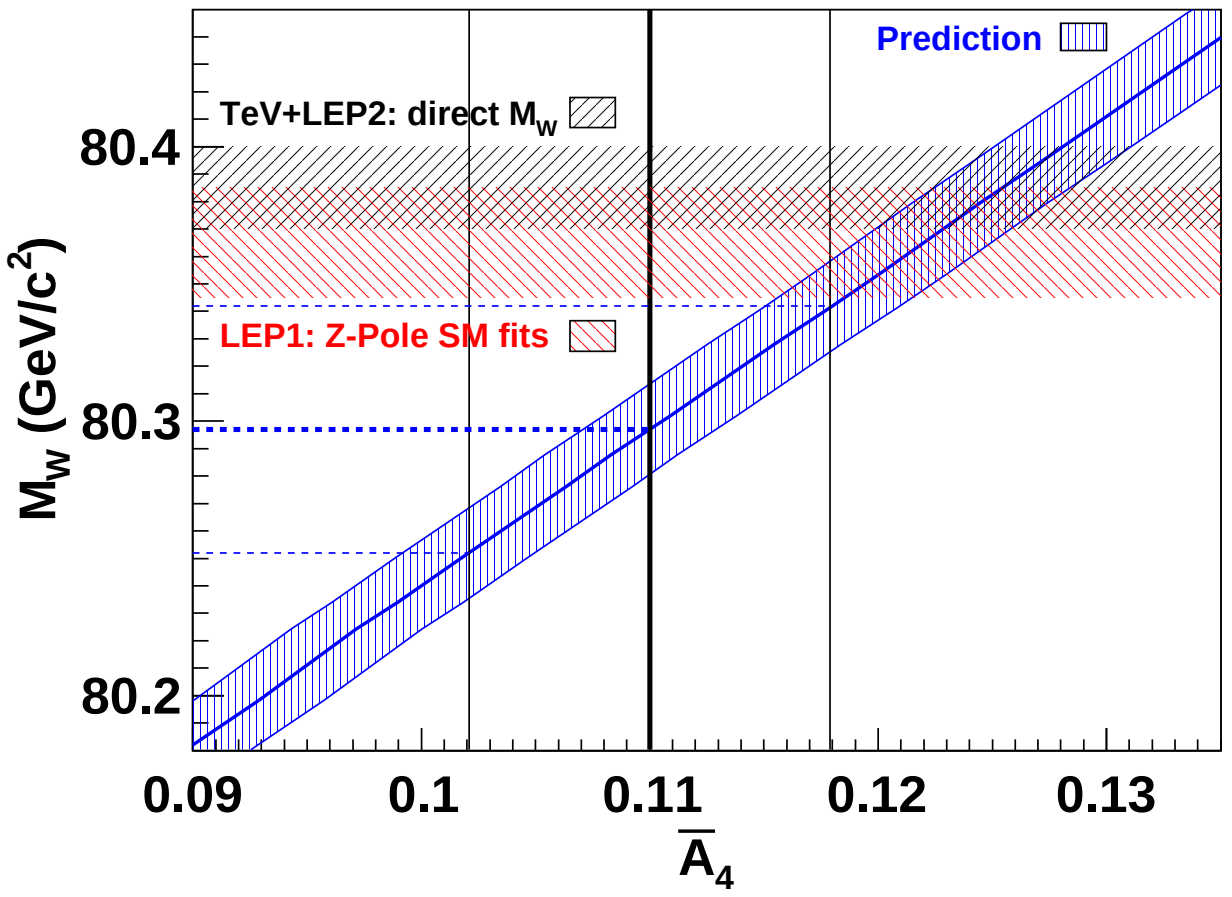

FIG. 13. Distribution of $M_W$ as a function of the $\bar{A}_4$ value as predicted by RESBOS. The prediction is the solid (blue) diagonal line and its one standard-deviation limits are the bands. The $\bar{A}_4$ measurement is the bold vertical line, and its one standard-deviation limits are the lighter vertical lines. The hatched horizontal bands are uncertainty limits from other $W$-mass measurements (see text).

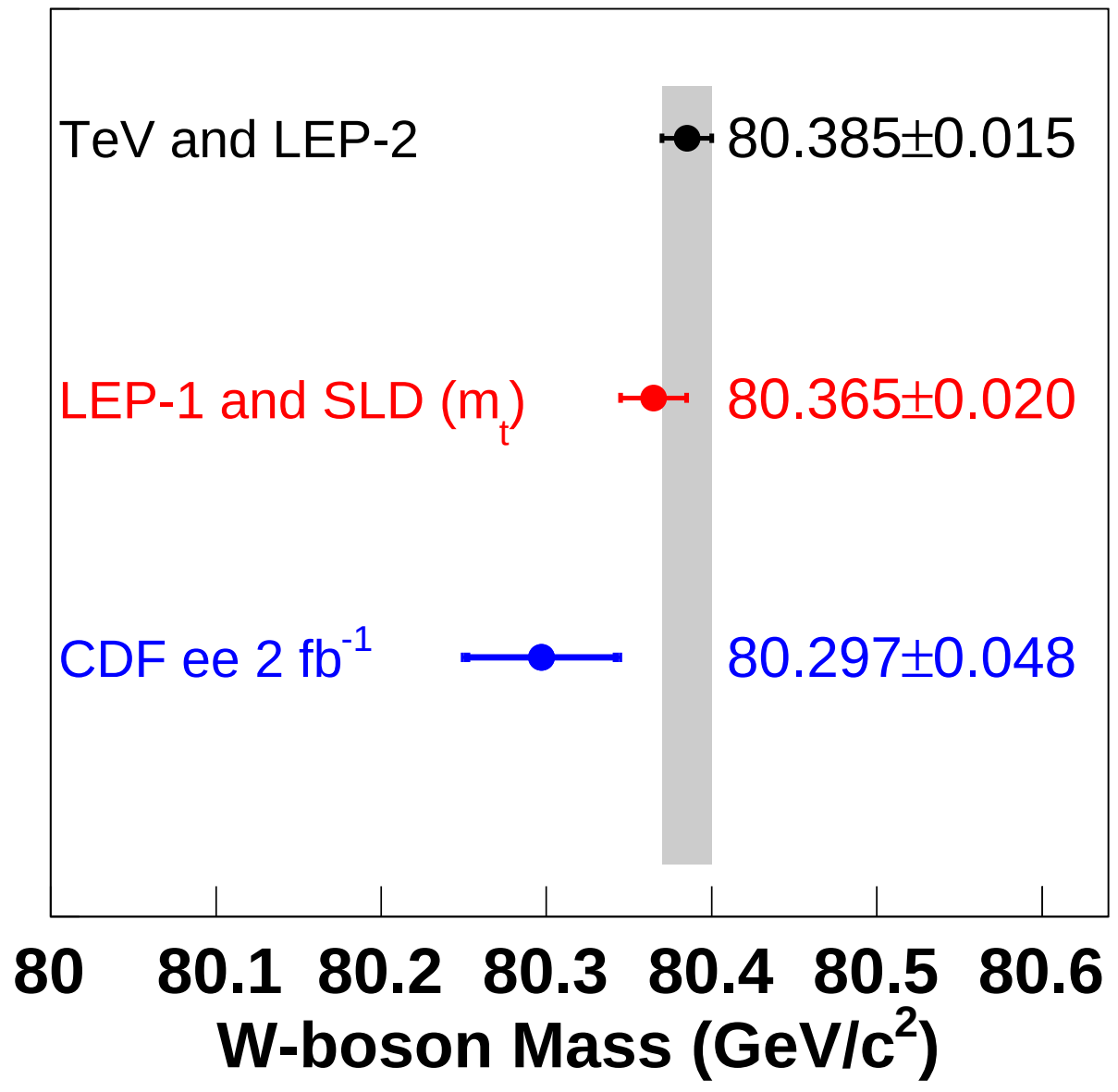

FIG. 14. Comparisons of experimental measurements of the $W$-boson mass: "TeV and LEP-2" represents direct measurements of the $W$-boson mass; "LEP-1 and SLD ($m_t$)" represents the standard-model analysis of $Z$-pole measurements; and "CDF ee 2 fb$^{-1}$" represents this analysis. The horizontal bars represent total uncertainties. For this analysis, the inner uncertainty bar is the measurement uncertainty.

The $\sin^2\theta_W$ parameter also specifies the correspondence between the $A_4$ angular coefficient and the effective-mixing parameters. As the parameters are averaged in the $\bar{A}_4$ angular coefficient, a reference value of the effective-leptonic mixing parameter at the $Z$ pole,

$$\sin^2\theta^{\rm lept}_{\rm eff} = {\rm Re}\,\kappa_e(s_Z, \sin^2\theta_W)\sin^2\theta_W,$$

is provided for comparisons. Although the $\bar{A}_4$ coefficient is integrated across the $\sqrt{s}$ range of 66–116 GeV, the bulk of the integrated cross section is near the vicinity of the $Z$ pole ($s_Z = M_Z^2$). Therefore, it is an effective probe of the leptonic $\sin^2\theta_{\rm eff}$ at the reference $s_Z$ value. The reference value of $\sin^2\theta^{\rm lept}_{\rm eff}$ corresponding to the $\bar{A}_4$ angular-coefficient measurement is

$$\sin^2\theta^{\rm lept}_{\rm eff} = 0.2328 \pm 0.0010,$$

where both statistical and systematic uncertanities are included. The experimental statistical uncertainty is $\pm 0.0009$. The systematic uncertainty, predominantly from the prediction, is $\pm 0.0003$. Relative to $\sin^2\theta^{\rm lept}_{\rm eff}$, the effective-mixing parameters of the $u$- and $d$-type quarks ${\rm Re}\,\kappa_{u,d}\sin^2\theta_W$ (at $s_Z$), are lower by 0.0001 and 0.0002, respectively. The corresponding $\sin^2\theta^{\rm lept}_{\rm eff}$ measurements from LEP-1 and SLD are

$$\begin{aligned} &0.23153 \pm 0.00016 \;\; (Z\text{-pole}) \;\text{ and} \\ &0.2320 \pm 0.0021 \;\; (\text{light quarks}), \end{aligned}$$

where the "$Z$-pole" measurement is from the standard-model analysis of the combined $Z$-pole results, and the "light quarks" measurement is from the light-quark ($u$, $d$, and $s$) asymmetries [5]. The previous corresponding Tevatron value from D0 derived from a measurement of $A_{\rm fb}(M)$ is $\sin^2\theta^{\rm lept}_{\rm eff} = 0.2309 \pm 0.0008 \pm 0.0006$, where the first contribution to the uncertainty is statistical and second systematic [4]. Figure 15 shows a comparison of these $\sin^2\theta^{\rm lept}_{\rm eff}$ measurements.

The admixture of light quarks in the Drell-Yan production and $e^+e^-$ collisions is somewhat different. The contributions of the various quarks to the incoming parton flux in Tevatron $p\bar{p}$ collisions are evaluated with the CT10 PDFs at a virtuality scale of $Q = M_Z$ and at a momentum fraction of $x = 0.047$ (corresponding to $\sqrt{s} = M_Z$). The $q\bar{q}$ fluxes of the $d$, $s$, $c$, and $b$ quarks relative to the $u$-quark flux are 0.51, 0.06, 0.02, and 0.01, respectively.

The EBA-based QCD calculations include the full electroweak radiative correction formalism of ZFITTER. Without this formalism, the extracted values of $\sin^2\theta^{\rm lept}_{\rm eff}$ tend to be slightly lower. For the value $\bar{A}_4 = 0.1100$, the difference between the derived value of $\sin^2\theta^{\rm lept}_{\rm eff}$ with and without the ZFITTER formalism for the RESBOS calculation is 0.0002. The corresponding value for the POWHEG-BOX calculation is 0.0003. The difference between the EBA-based RESBOS value and the non-EBA PYTHIA 6.41 value obtained with CTEQ5L PDFs is 0.0005. These differences are not negligible for precision measurements.

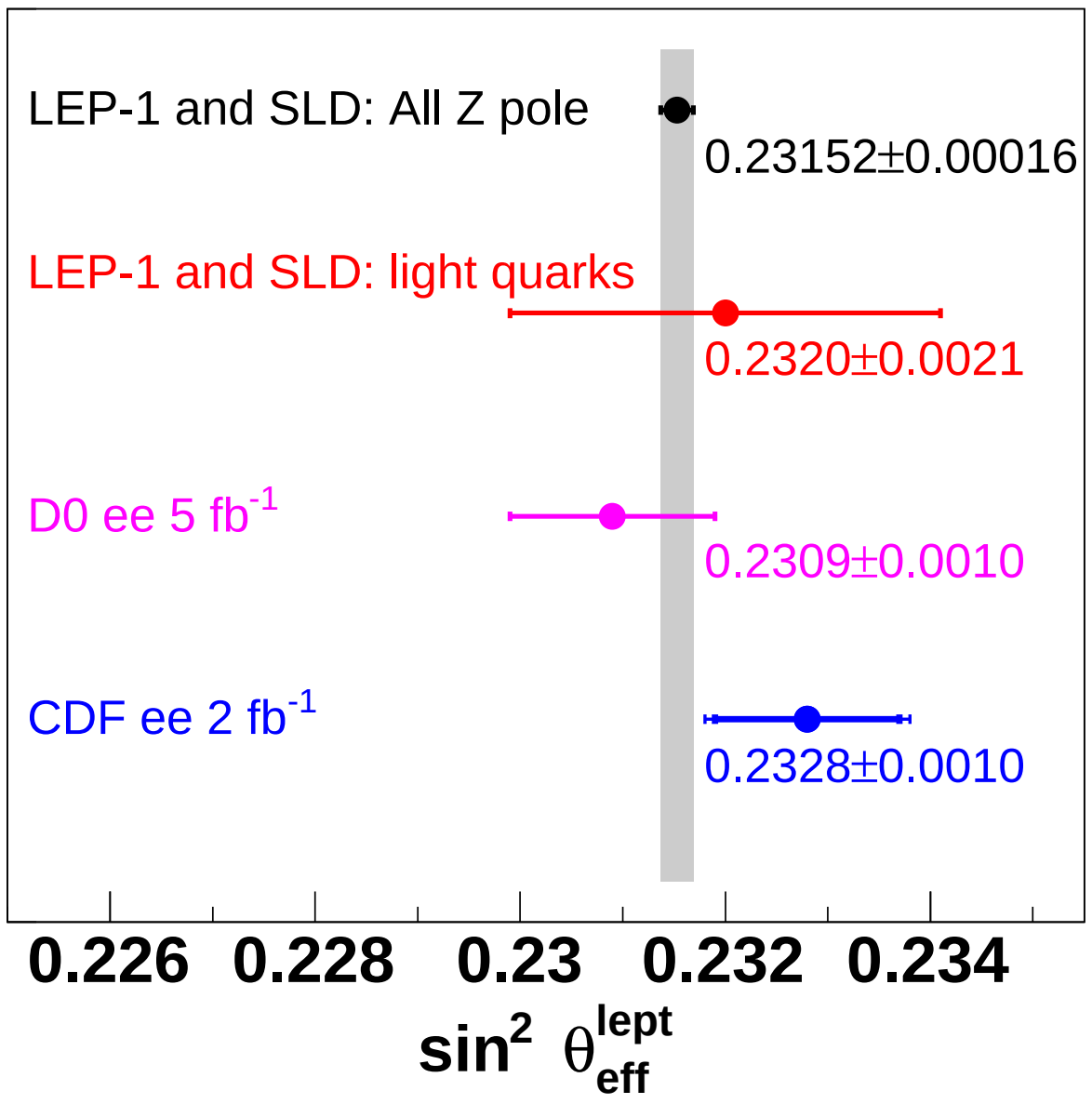

FIG. 15. Comparisons of experimental measurements of $\sin^2\theta^{\rm lept}_{\rm eff}$: "All $Z$ pole" represents the LEP-1 and SLD standard-model analysis of $Z$-pole measurements and "light quarks" represents the LEP-1 and SLD results from the light-quark asymmetries; "D0 ee 5 fb$^{-1}$" represents the D0 $A_{\rm fb}(M)$ analysis; and "CDF ee 2 fb$^{-1}$" represents this analysis. The horizontal bars represent total uncertainties. For this analysis, the inner uncertainty bar is the measurement uncertainty.

## VII. SUMMARY

The angular distribution of Drell-Yan $e^+e^-$ pairs provides information on the electroweak-mixing parameter $\sin^2\theta_W$. The electron forward-backward asymmetry in the polar-angle distribution $\cos\vartheta$ is governed by the $A_4\cos\vartheta$ term, whose $A_4$ coefficient is directly related to the $\sin^2\theta^{\rm lept}_{\rm eff}$ mixing parameter at the lepton vertex, and indirectly to $\sin^2\theta_W$. A new method for the determination of $\sin^2\theta^{\rm lept}_{\rm eff}$ using the average value of $A_4$ ($\bar{A}_4$) for $ee$-pairs in the $Z$-boson mass region of 66–116 GeV/$c^2$ is tested. The method utilizes standard-model calculations of $\bar{A}_4$ for different input values of $\sin^2\theta_W$, or equivalently, $\sin^2\theta^{\rm lept}_{\rm eff}$, for comparison with the measured value of $\bar{A}_4$. These calculations include both quantum chromodynamic and electroweak radiative corrections. The result for $\sin^2\theta_W$ is equivalent to an indirect determination of the $W$-boson mass. However, unlike $\sin^2\theta^{\rm lept}_{\rm eff}$, the interpretation of $\sin^2\theta_W$ or the $W$-boson mass is dependent on the standard-model context. Using the value $\bar{A}_4 = 0.1100 \pm 0.0079$ observed in a sample corresponding to 2.1 fb$^{-1}$ of integrated luminosity from $p\bar{p}$ collisions at a center-of-momentum energy of 1.96 TeV,

$$\sin^2\theta^{\rm lept}_{\rm eff} = 0.2328 \pm 0.0010,$$
$$\sin^2\theta_W = 0.2246 \pm 0.0009, \text{ and}$$
$$M_W(\text{indirect}) = 80.297 \pm 0.048 \text{ GeV}/c^2 .$$

Each uncertainty includes statistical and systematic contributions. Both results are consistent with LEP-1 and SLD $Z$-pole measurements. The value of $\sin^2\theta^{\rm lept}_{\rm eff}$ is also consistent with the previous Tevatron value from D0. The results of the test for the new method are promising. As the uncertainties are predominantly statistical, the measurement will improve with the analysis of the full Tevatron sample corresponding to 9 fb$^{-1}$ of integrated luminosity.

## ACKNOWLEDGMENTS

We thank T. Riemann for useful discussions and help on ZFITTER. We thank D. Wackeroth for useful discussions and help on the ZGRAD2 calculation. We thank the Fermilab staff and the technical staffs of the participating institutions for their vital contributions. This work was supported by the U.S. Department of Energy and National Science Foundation; the Italian Istituto Nazionale di Fisica Nucleare; the Ministry of Education, Culture, Sports, Science and Technology of Japan; the Natural Sciences and Engineering Research Council of Canada; the National Science Council of the Republic of China; the Swiss National Science Foundation; the A.P. Sloan Foundation; the Bundesministerium für Bildung und Forschung, Germany; the Korean World Class University Program, the National Research Foundation of Korea; the Science and Technology Facilities Council and the Royal Society, UK; the Russian Foundation for Basic Research; the Ministerio de Ciencia e Innovación, and Programa Consolider-Ingenio 2010, Spain; the Slovak R&D Agency; the Academy of Finland; the Australian Research Council (ARC); and the EU community Marie Curie Fellowship contract 302103.

### Appendix A: ZFITTER

The input parameters to the ZFITTER radiative-correction calculation are particle masses, the electromagnetic fine-structure constant $\alpha_{em}$, the Fermi constant $G_F$, the strong coupling at the $Z$ mass $\alpha_s(M_Z^2)$, and the contribution of the light quarks to the "running" $\alpha_{em}$ at the $Z$ mass $\Delta\alpha^{(5)}_{em}(M_Z^2)$ (DALH5). The scale-dependent couplings are $\alpha_s(M_Z^2) = 0.118$ and $\Delta\alpha^{(5)}_{em}(M_Z^2) = 0.0275$ [33]. The mass parameters are $M_Z = 91.1875$ GeV/$c^2$ [5], $m_t = 173.2$ GeV/$c^2$ (top quark) [34], and $m_H = 125$ GeV/$c^2$ (Higgs boson). Form factors and the $Z$-boson total-decay width $\Gamma_Z$, are calculated.

The renormalization scheme used by ZFITTER is the on-shell scheme [10], where particle masses are on-shell, and

$$\sin^2\theta_W = 1 - M_W^2/M_Z^2$$

holds to all orders of perturbation theory by definition. If both $G_F$ and $m_H$ are specified, $\sin\theta_W$ is not independent, and is derived from standard-model constraints that use radiative corrections. To vary the $\sin\theta_W$ ($M_W$) parameter, the value of $G_F$ is changed by a small amount prior to the calculation so that the derived $M_W$ range is 80.0–80.5 GeV/$c^2$.[1] The set of $M_W$ values corresponds to a family of physics models with standard-model like couplings where $\sin^2\theta_W$ and the coupling ($G_F$) are defined by the $M_W$ parameter. The Higgs-boson mass constraint $m_H = 125$ GeV/$c^2$ keeps the form factors within the vicinity of standard-model fit values from LEP-1 and SLD [5]. The primary purpose of ZFITTER is to provide tables of form factors for each model.

Access to ZFITTER calculations is through its *interfaces.* The calculation of form factors uses ZFITTER's interface to its $e\bar{e} \to Z \to f\bar{f}$ scattering-amplitude formalism (ROKANC). External QED and QCD radiation are turned off. The form factors include corrections from $\gamma$-$Z$ mixing effects and from non-resonant $\gamma$ and $Z$ exchanges. The contributions from $WW$ and $ZZ$ box diagrams are included, but as they are not multiplicative form-factor corrections, these corrections are only approximate. The calculation is done in the massless-fermion approximation so the form factors only depend on the fermion weak isospin and charge. Consequently, the form factors are distinguished via three indices: $e$ (electron type), $u$ (up-quark type), and $d$ (down-quark type). The form factors are functions of the Mandelstam variable $s$, and with the inclusion of the box diagrams they also depend on $t = -\frac{1}{2}s(1-\cos\theta)$, where $\theta$ is the angle between the $e$ and $f$. The ZFITTER scattering-amplitude ansatz is

$$\begin{aligned} A_q = \frac{i}{4}\, & \frac{\sqrt{2}G_F M_Z^2}{s - (M_Z^2 - i\, s\Gamma_Z/M_Z)}\, 4T_3^e T_3^q\, \rho_{eq} \\ & [\langle\bar{e}|\gamma^\mu(1+\gamma_5)|e\rangle\langle\bar{q}|\gamma_\mu(1+\gamma_5)|q\rangle + \\ & -4|Q_e|\kappa_e \sin^2\theta_W\, \langle\bar{e}|\gamma^\mu|e\rangle\langle\bar{q}|\gamma_\mu(1+\gamma_5)|q\rangle + \\ & -4|Q_q|\kappa_q \sin^2\theta_W\, \langle\bar{e}|\gamma^\mu(1+\gamma_5)|e\rangle\langle\bar{q}|\gamma_\mu|q\rangle + \\ & 16|Q_e Q_q|\kappa_{eq}\sin^4\theta_W \langle\bar{e}|\gamma^\mu|e\rangle\langle\bar{q}|\gamma_\mu|q\rangle]\,, \end{aligned} \quad \text{(A1)}$$

where $q = u$ or $d$, the $\rho_{eq}$, $\kappa_e$, $\kappa_q$, and $\kappa_{eq}$ are complex-valued form factors, the bilinear $\gamma$ matrix terms covariantly are contracted, and $\frac{1}{2}(1+\gamma_5)$ is the left-handed helicity projector in the ZFITTER convention. The $\rho_{eq}$ form factor is defined to be used with $G_F$. As their significant radiative corrections cancel to a large extent, they are combined to minimize the size of applied corrections. At $s = M_Z^2$, the $\kappa_e$ form factors of the $A_u$ and $A_d$ amplitudes are numerically the same.

The amplitude $A_q$ can be approximated with these Born-level $g_V^f$ and $g_A^f$ replacements,

$$\begin{aligned} g_V^f &\to \sqrt{\rho_{eq}}\,(T_3^f - 2Q_f\kappa_f\, \sin^2\theta_W) \\ g_A^f &\to \sqrt{\rho_{eq}}\,T_3^f\,, \end{aligned}$$

where $f = e$ or $q$. The Born electron-quark current-current amplitude is nearly identical to $A_q$ except that the last term contains $\kappa_e\kappa_q\sin^4\theta_W$ rather than $\kappa_{eq}\sin^4\theta_W$. The $\kappa_{eq}$ form factor must be explicitly incorporated into the Born amplitude for a full implementation of the ZFITTER $A_q$ amplitude; this is accomplished with the addition of an amplitude-correction term containing the $\kappa_{eq} - \kappa_e\kappa_q$ form factor. The space-time structure of the amplitude for the photon and the $\kappa_{eq} - \kappa_e\kappa_q$ correction is identical, and their amplitudes may be consolidated into a single term.

The $s$ and $t$ ($\cos\theta$) dependencies of the form factors are illustrated for $\sin^2\theta_W = 0.2231$ in distributions of the form factor as a function of $\cos\theta$, where curves of different $s$ are superimposed on the same panel. The range of $s$ is $66 < \sqrt{s} < 116$ GeV, and displayed in 5 GeV intervals. The real parts of the form factors $\rho_{eq}$, $\kappa_e$, $\kappa_q$, and $\kappa_{eq}$ are shown in Figs. 16, 17, 18, and 19, respectively. The imaginary part of these form factors is on the order of $\pm 0.02$ in value.

The $t$ variation (from the box diagrams) for each $s$ is averaged out, and this average is a cross-section (Born $d\sigma/d\cos\theta$) weighted average. The form factors used in QCD calculations are implemented as complex-valued look-up tables in ($\sin^2\theta_W$, $s$).

Only the photon self-energy correction from fermion loops is used with the ZFITTER $Z$-amplitude form factors. The correction is applied as a form factor to the photon propagator

$$\frac{i\, e^2 Q_e Q_q}{s} \to \frac{i\, e^2 Q_e Q_q}{s} \frac{1}{1 - \Delta\alpha_{em}(s)}\,,$$

where $1 - \Delta\alpha_{em}(s)$ is the complex-valued form factor, which equals 1 when $s = 0$. The fermion-loop integrals of the form factor are complex-valued functions of $s$ and the fermion mass, $m_f$. All fermion pairs above production thresholds, *i.e.*, $4m_f^2 < s$ contribute to the imaginary part of the form factor. The leptonic-loop contributions and the imaginary part of quark loops are calculated. The contribution of the light quarks to the real part of the form factor is derived from measurements of $e^+e^- \to$ hadrons and is a function of $s$. At the $Z$ pole, the sum of contributions from the $u$, $c$, $d$, $s$, and $b$ quarks is $\Delta\alpha_{em}^{(5)}(M_Z^2) = 0.0275 \pm 0.0001$ [33]. Figure 20 illustrates $\Delta\alpha_{em}(s)$.

[1] The ZFITTER electroweak radiative correction package (DIZET) is first used to iteratively estimate $G_F$ from a target $M_W$ input (IMOMS=3). Form factors are not calculated due to a partial implementation. The code which calculates constants (CONST1) is modified to use this new $G_F$, then form factors are calculated using the default method (DIZET with IMOMS=1).

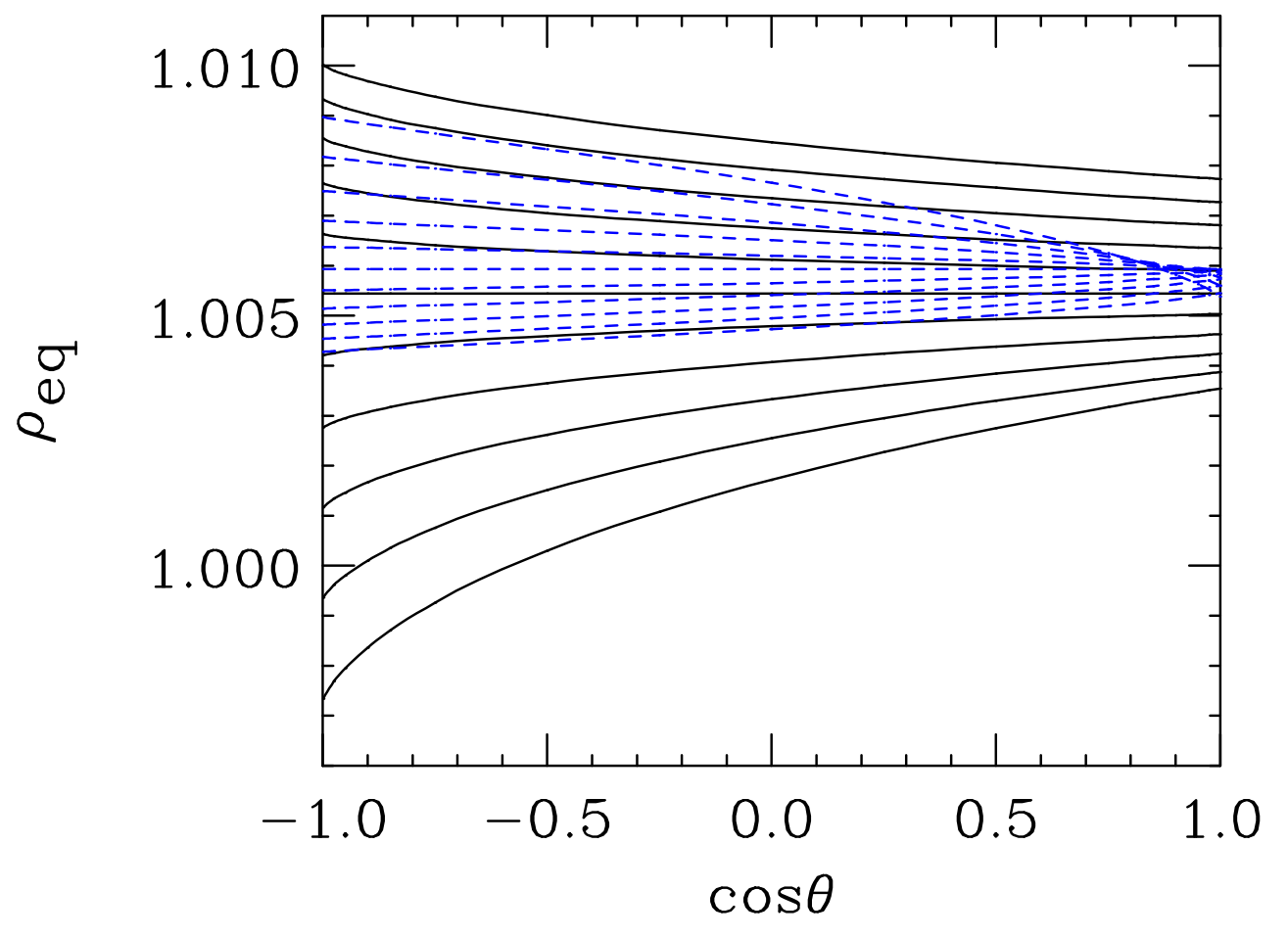

FIG. 16. Real part of $\rho_{eq}$ as a function of $\cos\theta$ for $\sin^2\theta_W = 0.2231$. Each curve corresponds to a different value of $\sqrt{s}$, varying from 66 to 116 GeV. The curves change monotonically with each step of $s$. The solid (black) curves are for $u$-type amplitudes, and the dashed (blue) curves are for $d$-type amplitudes. For the $u$-type amplitude, the highest mass corresponds to the lowermost curve at $\cos\theta = -1$, and for the $d$-type amplitude, the highest mass corresponds to the uppermost curve at $\cos\theta = -1$. The flat lines in the middle correspond to $\sqrt{s} = M_Z$.

### Appendix B: EBA Operational Tests

The ZGRAD2 calculation [23] is a LO QCD calculation with $\mathcal{O}(\alpha)$ standard-model corrections to the Drell-Yan $p\bar{p} \to e^+e^-$ process. As the calculation of EWK corrections differs from that of ZFITTER, it provides a test of the implementation of the ZFITTER form-factor input to the EBA calculations. A full test is not possible because a few parts of the ZFITTER EBA implementation differ from ZGRAD2. Form-factor corrections are calculated by ZGRAD2 for the $g_A^f$ and $g_V^f$ couplings of both the $\gamma$ and $Z$ bosons, *i.e.*, $g_{A,V}^f \to F_{A,V}^f\, g_{A,V}^f$, where $F_{A,V}^f$ is the form factor. Bosonic self-energy corrections are included. In the cross-section amplitude, the corrected $g_A^f$ and $g_V^f$ are complex-valued couplings. The $WW$ and $ZZ$ box diagram cross-sections are separately calculated, and added to the total cross section. For the following test, both box-diagram and initial- and final-state QED radiation contributions are disabled. The couplings from ZGRAD2 are converted into ZFITTER ($\rho$ and $\kappa$) form factors, and the ratio of the ZGRAD2-to-ZFITTER form factors (which are complex valued) are evaluated for comparisons. The $\kappa$ form factors are very similar for $\sin^2\theta_W = 0.2230$: The fractional differences of both the real and imaginary parts of the ratio range from −0.1% to 0.2% over $66 < \sqrt{s} < 116$ GeV. The $\rho$ form factors have offsets over the range of $\sqrt{s}$. The real part decreases from −0.5% to −0.7%, and the imaginary part increases from 0.2% to 0.5%. The $Z$-boson coupling schemes of ZGRAD2 and ZFITTER differ, and can affect $\rho$.

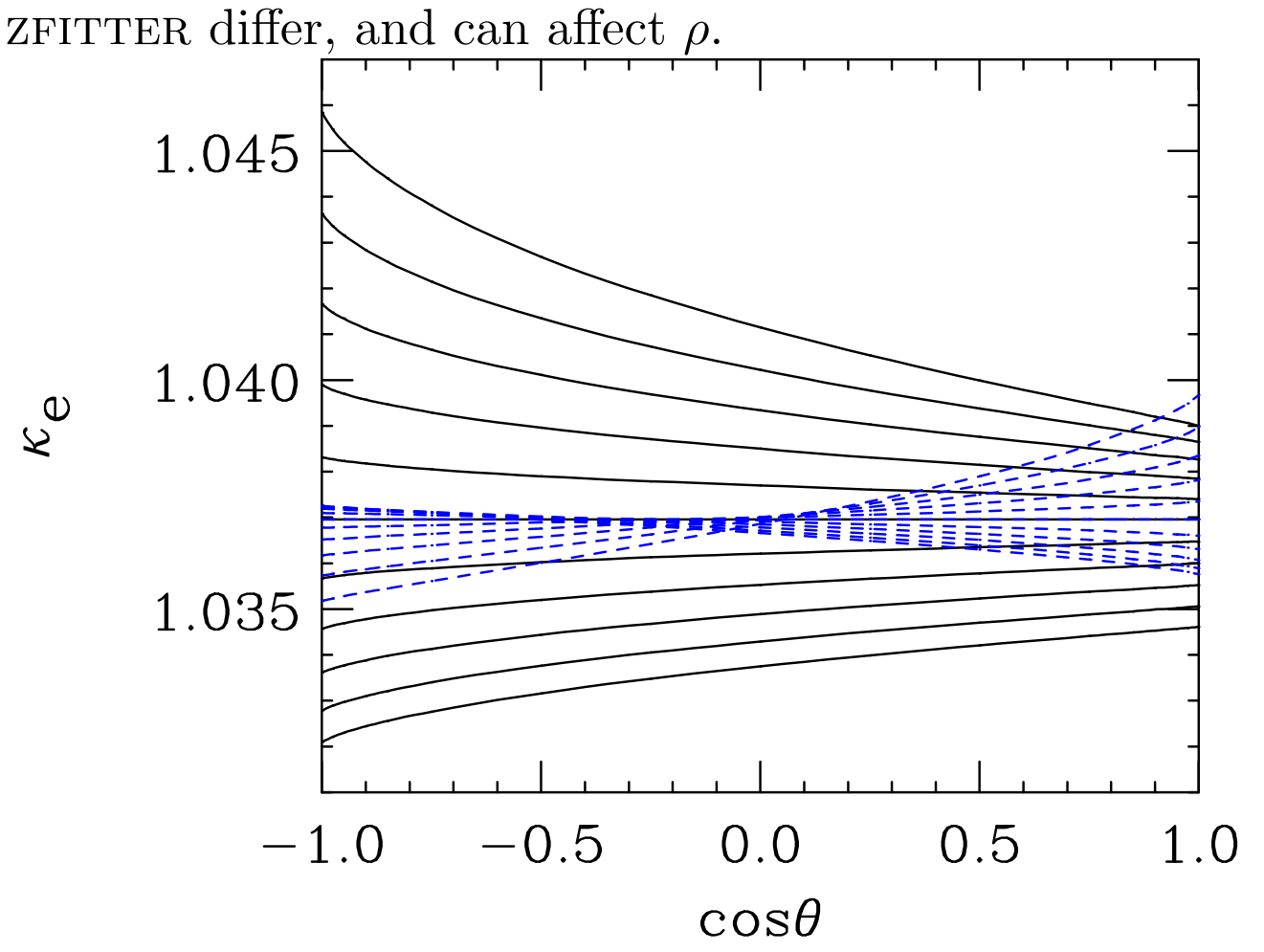

FIG. 17. Real part of $\kappa_e$ as a function of $\cos\theta$ for $\sin^2\theta_W = 0.2231$. Each curve corresponds to a different value of $\sqrt{s}$, varying from 66 to 116 GeV. The curves change monotonically with each step of $s$. The solid (black) curves are for $u$-type amplitudes, and the dashed (blue) curves are for $d$-type amplitudes. For the $u$-type amplitude, the highest mass corresponds to the uppermost curve at $\cos\theta = -1$, and for the $d$-type amplitude, the highest mass corresponds to the lowermost curve at $\cos\theta = -1$. The flat lines in the middle corresponds to $\sqrt{s} = M_Z$.

Next, the effect of $WW$ and $ZZ$ box diagrams on the value of the $\bar{A}_4$ coefficient is calculated with both the ZGRAD2 and the ZFITTER EBA-based tree calculation. For both, the effect is small and essentially the same: The value of the coefficient with box-diagram contributions is 0.0001 smaller in difference than without box-diagram contributions. This confirms that the averaging of the $t$ dependence of the ZFITTER form factors from the box diagrams used in the EBA form-factor tables does not impact the EBA-based calculations.

In standard-model tests of the process $e^+e^- \to f\bar{f}$, ZFITTER calculates cross sections and final-state fermion asymmetries using all form factors in their complex-valued form: the vertex form factors $\rho_{eq}$, $\kappa_e$, $\kappa_q$, and $\kappa_{eq}$, and the photon self-energy correction form factor. The ZGRAD2 calculations do not have the $\kappa_{eq}$ form factor or the imaginary part of the photon self-energy correction form factor. These corrections, along with the difference in the $\rho$ form factor, induce a shift of −0.0025 in the value of $\bar{A}_4$ from the default EBA-based tree calculation, with 75% due to the imaginary part of the photon self-energy correction. The calculation of $\bar{A}_4$ by ZGRAD2 yields a value $0.0036 \pm 0.0006$ smaller than the ZFITTER EBA-based tree calculation, but is consistent with the expected difference.

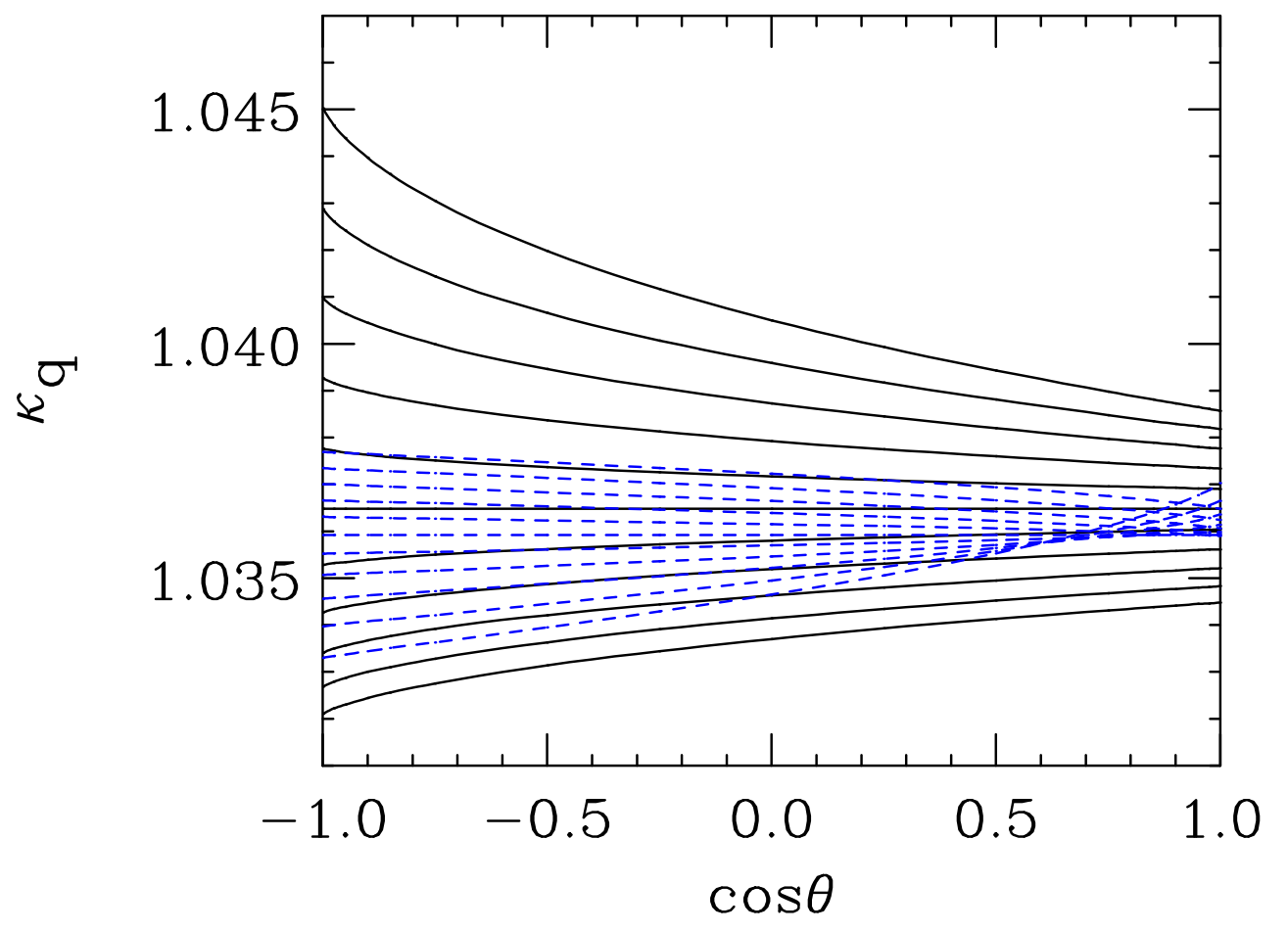

FIG. 18. Real part of $\kappa_q$ as a function of $\cos\theta$ for $\sin^2\theta_W = 0.2231$. Each curve corresponds to a different value of $\sqrt{s}$, varying from 66 to 116 GeV. The curves change monotonically with each step of $s$. The solid (black) curves are for $u$-type amplitudes, and the dashed (blue) curves are for $d$-type amplitudes. For the $u$-type amplitude, the highest mass corresponds to the uppermost curve at $\cos\theta = -1$, and for the $d$-type amplitude, the highest mass corresponds to the lowermost curve at $\cos\theta = -1$. The flat lines in the middle corresponds to $\sqrt{s} = M_Z$.

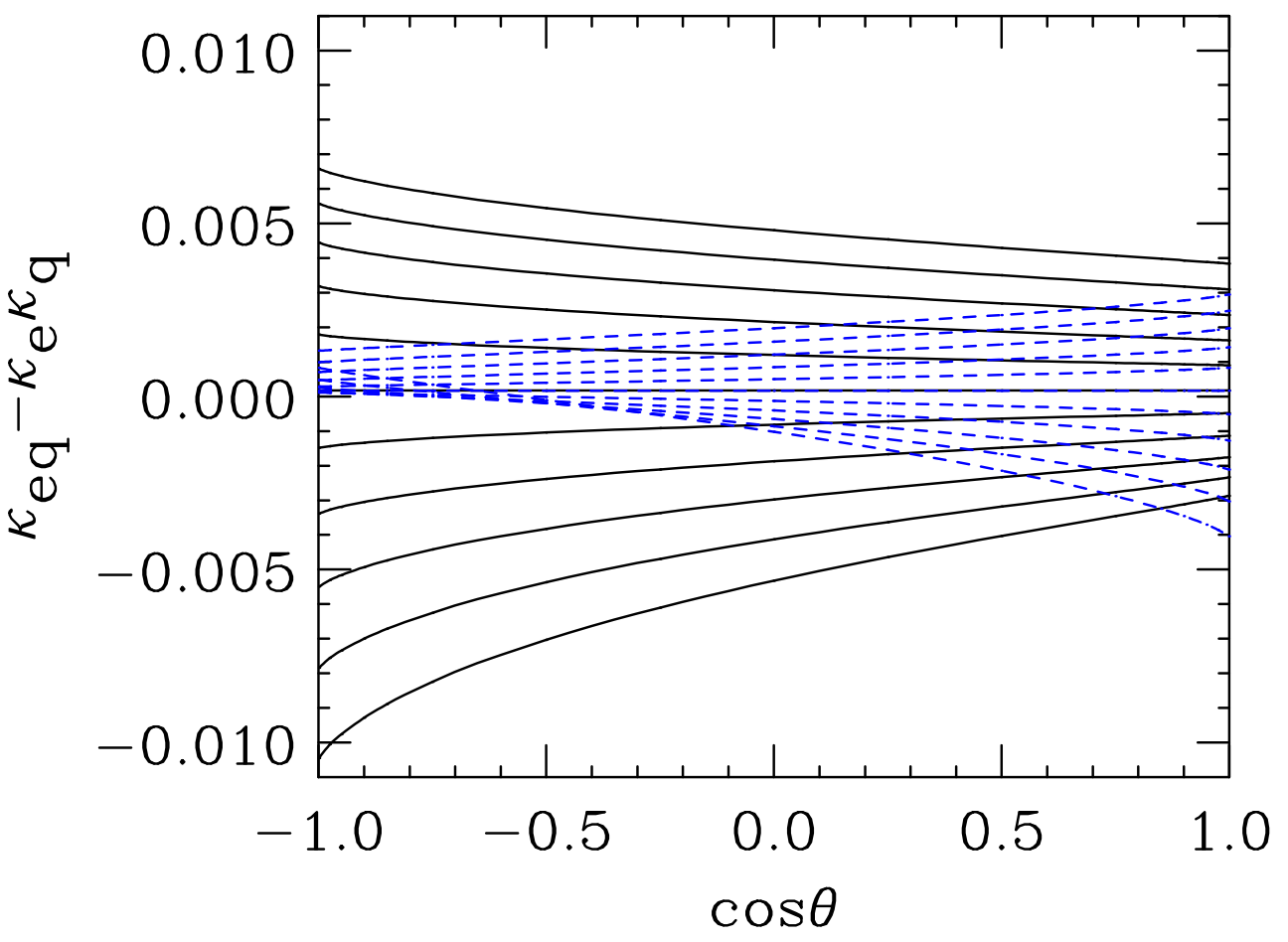

FIG. 19. Real part of $\kappa_{eq} - \kappa_e \kappa_q$ as a function of $\cos\theta$ for $\sin^2\theta_W = 0.2231$. Each curve corresponds to a different value of $\sqrt{s}$, varying from 66 to 116 GeV. The curves change monotonically with each step of $s$. The solid (black) curves are for $u$-type amplitudes, and the dashed (blue) curves are for $d$-type amplitudes. For the $u$-type amplitude, the highest mass corresponds to the lowermost curve at $\cos\theta = -1$, and for the $d$-type amplitude, the highest mass corresponds to the uppermost curve at $\cos\theta = -1$. The flat lines in the middle correspond to $\sqrt{s} = M_Z$.

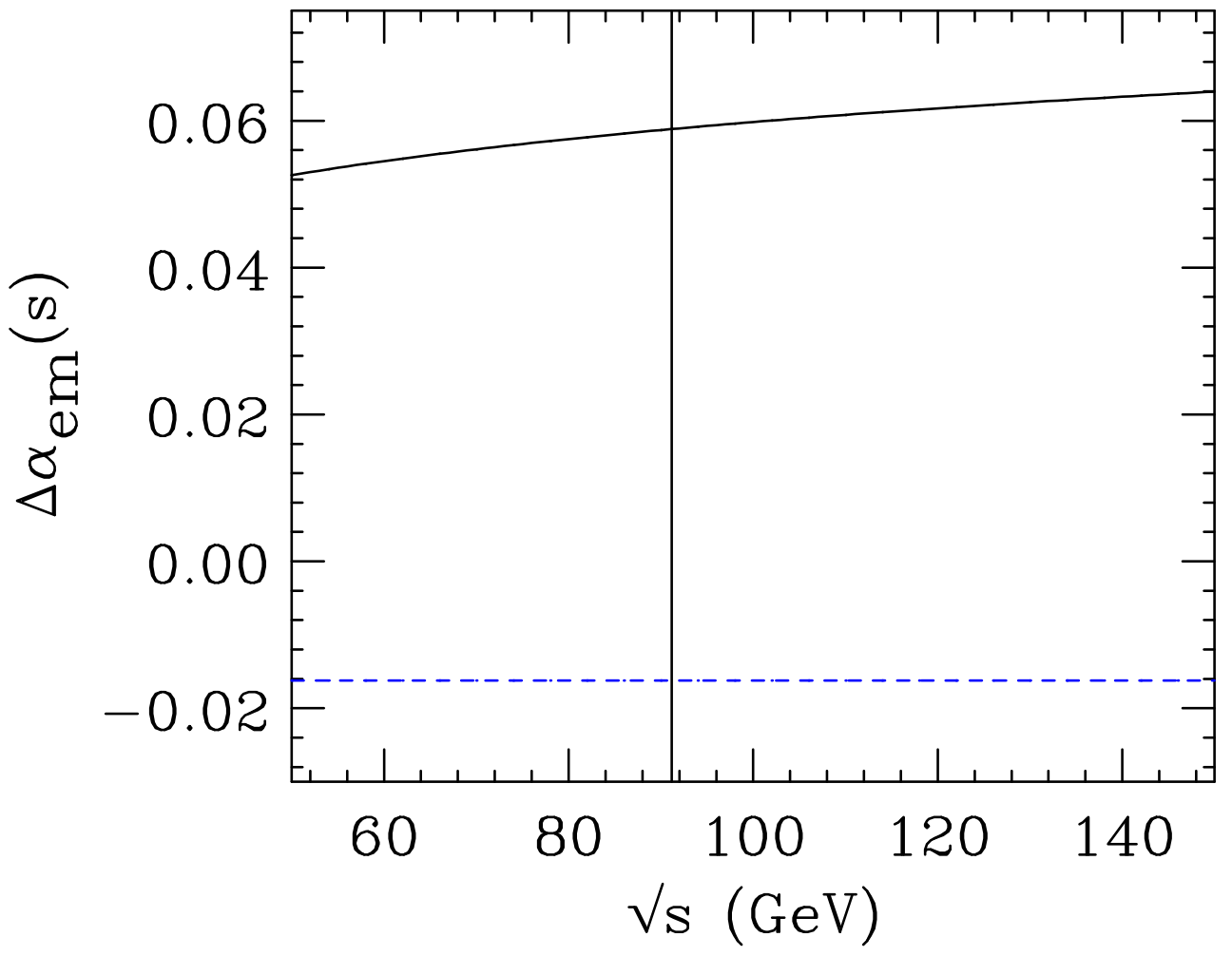

FIG. 20. The $\Delta\alpha_{em}(s)$ form factor for $50 < \sqrt{s} < 150$ GeV. The upper solid curve corresponds to the real part and the lower dashed curve corresponds the imaginary part. The vertical line is $\sqrt{s} = M_Z$.